\documentclass[twoside]{IEEEtran}

\usepackage{cite}

\usepackage{amsmath,amssymb,amsfonts,mathrsfs,bm}

\DeclareSymbolFont{tnptimes}{OT1}{ptm}{m}{n}
\DeclareMathSymbol{0}{\mathalpha}{tnptimes}{48}
\DeclareMathSymbol{1}{\mathalpha}{tnptimes}{"31}
\DeclareMathSymbol{2}{\mathalpha}{tnptimes}{"32}
\DeclareMathSymbol{3}{\mathalpha}{tnptimes}{"33}
\DeclareMathSymbol{4}{\mathalpha}{tnptimes}{"34}
\DeclareMathSymbol{5}{\mathalpha}{tnptimes}{"35}
\DeclareMathSymbol{6}{\mathalpha}{tnptimes}{"36}
\DeclareMathSymbol{7}{\mathalpha}{tnptimes}{"37}
\DeclareMathSymbol{8}{\mathalpha}{tnptimes}{"38}
\DeclareMathSymbol{9}{\mathalpha}{tnptimes}{"39}

\usepackage{graphicx}
\usepackage{subfigure}
\graphicspath{{figures/}}

\usepackage{tabularx}
\usepackage{booktabs}

\usepackage{verbatim}

\makeatletter
  \newif\if@restonecol
  \def\@IEEEBIOskipN{5mm}
\makeatother

\usepackage[linesnumbered,ruled,lined]{algorithm2e}
\usepackage{algpseudocode}

\usepackage{flushend}

\usepackage{scalerel}
\usepackage{tikz}
\usetikzlibrary{svg.path}

\definecolor{orcidlogocol}{HTML}{A6CE39}
\tikzset{
  orcidlogo/.pic={
    \fill[orcidlogocol] svg{M256,128c0,70.7-57.3,128-128,128C57.3,256,0,198.7,0,128C0,57.3,57.3,0,128,0C198.7,0,256,57.3,256,128z};
    \fill[white] svg{M86.3,186.2H70.9V79.1h15.4v48.4V186.2z}
                 svg{M108.9,79.1h41.6c39.6,0,57,28.3,57,53.6c0,27.5-21.5,53.6-56.8,53.6h-41.8V79.1z M124.3,172.4h24.5c34.9,0,42.9-26.5,42.9-39.7c0-21.5-13.7-39.7-43.7-39.7h-23.7V172.4z}
                 svg{M88.7,56.8c0,5.5-4.5,10.1-10.1,10.1c-5.6,0-10.1-4.6-10.1-10.1c0-5.6,4.5-10.1,10.1-10.1C84.2,46.7,88.7,51.3,88.7,56.8z};
  }
}

\newcommand\orcidicon[1]{%
  \href{https://orcid.org/#1}{
    \mbox{\scalerel*{
      \begin{tikzpicture}[yscale=-1, transform shape]
        \pic{orcidlogo};
      \end{tikzpicture}
    }{|}%
    }%
  }%
}

\newlength{\subfiglen}

\usepackage[colorlinks,allcolors=black]{hyperref} 

\newtheorem{theorem}{Theorem}
\newtheorem{lemma}{Lemma}

\newtheorem{remark}{Remark}

\newtheorem{assumption}{Assumption}

\DeclareMathOperator*{\argmax}{argmax}

\title{Adaptive Testing Environment Generation for Connected and Automated Vehicles with\\ Dense Reinforcement Learning}

\author{Jingxuan Yang\textsuperscript{\orcidicon{0000-0001-9798-7347}}, Ruoxuan Bai\textsuperscript{\orcidicon{0009-0006-9290-004X}}, Haoyuan Ji\textsuperscript{\orcidicon{0009-0000-2177-8047}}, Yi Zhang\textsuperscript{\orcidicon{0000-0001-5526-866X}}, \IEEEmembership{Senior Member,~IEEE}, \\ Jianming Hu\textsuperscript{\orcidicon{0000-0001-8065-7309}}, \IEEEmembership{Senior Member,~IEEE}, and Shuo Feng\textsuperscript{\orcidicon{0000-0002-2117-4427}}, \IEEEmembership{Member,~IEEE}%
\thanks{This work was supported in part by Beijing Nova Program under Grant 20230484259 and in part by Beijing Natural Science Foundation 4244092. \textit{(Corresponding author: Shuo Feng.)}}%
\thanks{Jingxuan Yang, Ruoxuan Bai, Haoyuan Ji and Jianming Hu are with the Department of Automation, Tsinghua University, Beijing 100084, China (email: yangjx20@mails.tsinghua.edu.cn, bairx22@mails.tsinghua.edu.cn, jihy21@mails.tsinghua.edu.cn, hujm@mail.tsinghua.edu.cn).}%
\thanks{Yi Zhang is with the Department of Automation, Beijing National Research Center for Information Science and Technology (BNRist), Tsinghua University, Beijing 100084, China, also with the Tsinghua-Berkeley Shenzhen Institute (TBSI), Shenzhen 518055, China, and also with the Jiangsu Province Collaborative Innovation Center of Modern Urban Traffic Technologies, Nanjing 210096, China (e-mail: zhyi@mail.tsinghua.edu.cn).}%
\thanks{Shuo Feng is with the Department of Automation, Beijing National Research Center for Information Science and Technology (BNRist), Tsinghua University, Beijing 100084, China (e-mail: fshuo@tsinghua.edu.cn).}
\thanks{Digital Object Identifier 10.1109/TITS.2024.0000000}
}

\IEEEpubid{0000-0000~\copyright~2024 IEEE}

\begin{document}

\maketitle

\begin{abstract}
  The assessment of safety performance plays a pivotal role in the development and deployment of connected and automated vehicles (CAVs). A common approach involves designing testing scenarios based on prior knowledge of CAVs (e.g., surrogate models), conducting tests in these scenarios, and subsequently evaluating CAVs' safety performances. However, substantial differences between CAVs and the prior knowledge can significantly diminish the evaluation efficiency. In response to this issue, existing studies predominantly concentrate on the adaptive design of testing scenarios during the CAV testing process. Yet, these methods have limitations in their applicability to high-dimensional scenarios. To overcome this challenge, we develop an adaptive testing environment that bolsters evaluation robustness by incorporating multiple surrogate models and optimizing the combination coefficients of these surrogate models to enhance evaluation efficiency. We formulate the optimization problem as a regression task utilizing quadratic programming. To efficiently obtain the regression target via reinforcement learning, we propose the dense reinforcement learning method and devise a new adaptive policy with high sample efficiency. Essentially, our approach centers on learning the values of critical scenes displaying substantial surrogate-to-real gaps. The effectiveness of our method is validated in high-dimensional overtaking scenarios, demonstrating that our approach achieves notable evaluation efficiency.
\end{abstract}

\begin{IEEEkeywords}
  Adaptive testing environment generation, connected and automated vehicles, dense reinforcement learning
\end{IEEEkeywords}

\section{Introduction}
\label{sec:introduction}

\begin{figure}[!t]
  \centering
  \includegraphics[width=8cm]{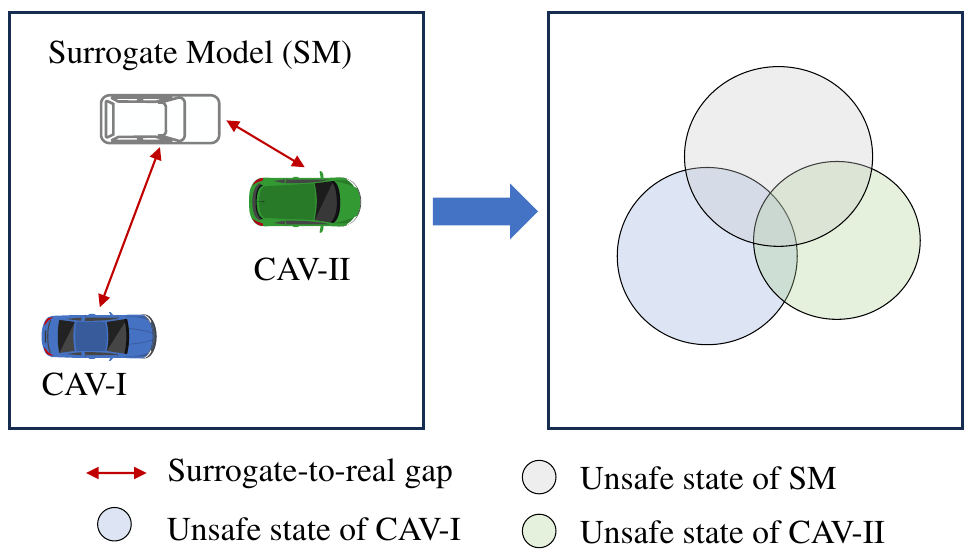}
  \caption{Illustration of the surrogate-to-real gaps.}
  \label{fig:surrogate_to_real_gap}
\end{figure}

\IEEEPARstart{T}{esting} and evaluating the safety performance of connected and automated vehicles presents notable challenges in their development and deployment. One suggested approach involves testing CAVs in the naturalistic driving environment (NDE), observing their behaviors, and statistically comparing the testing results with human drivers. However, the scarcity of safety-critical events in NDE necessitates an impractical amount of testing miles—sometimes in the hundreds of millions or even billions—to demonstrate CAVs' safety performance at the human-level, rendering the evaluation process intolerably inefficient \cite{kalra2016driving}. To increase evaluation efficiency, recent years have seen rapid advancements in the field of testing scenario library generation 
\cite{feng2023dense,yan2023learning,sun2021corner,li2024few,li2021scegene,wang2021advsim,menzel2018scenarios,tian2018deeptest,rempe2022generating,li2016intelligence,li2018artificial,li2019parallel,zhao2023genetic,riedmaier2020survey}.
This involves deliberately generating safety-critical testing scenarios using prior knowledge of CAVs, such as surrogate models (SMs). Employing SMs is promising for significantly enhancing the evaluation efficiency \cite{feng2021intelligent,feng2020parti,feng2020partii}. Nevertheless, due to the intricate nature and black-box characteristics of CAVs, substantial performance disparities exist between SMs and diverse CAVs, as shown in Fig.~\ref{fig:surrogate_to_real_gap}. This mismatch could undermine the effectiveness of the generated testing scenario libraries, ultimately diminishing the evaluation efficiency for diverse CAVs.

\IEEEpubidadjcol



\begin{figure}[!t]
  \centering
  \includegraphics[width=8cm]{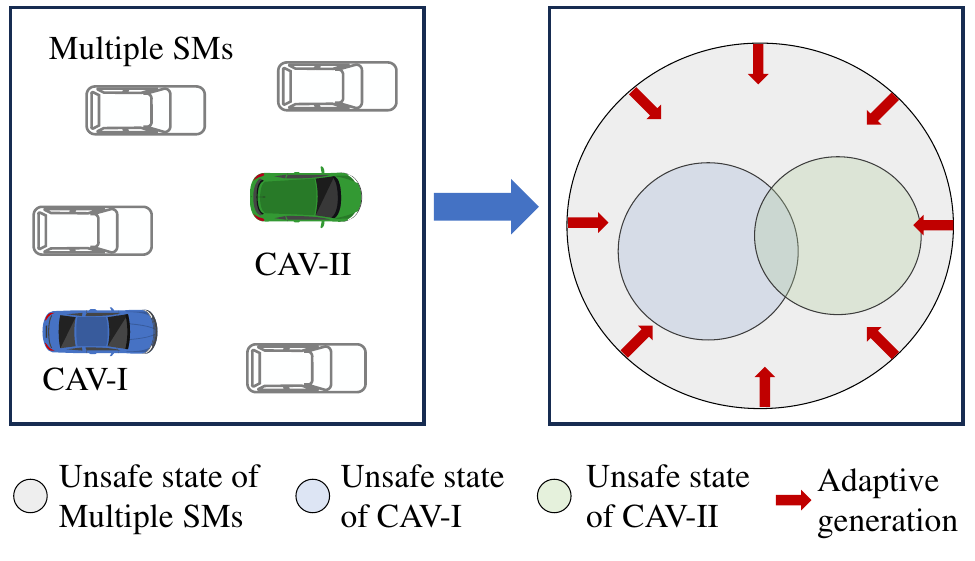}
  \caption{Illustration of the adaptive testing environment generation method with multiple SMs.}
  \label{fig:multiple_SMs}
\end{figure}



To tackle this issue, several adaptive testing methods have been proposed \cite{mullins2018adaptive,koren2018adaptive,feng2020adaptive,sun2021adaptive,yang2022adaptive,gong2023adaptive,yang2023adaptive}. The fundamental concept of these methods is to dynamically generate testing scenarios during the evaluation process of CAVs. As more testing results accumulate, more posterior knowledge of CAVs is gained, enabling the customization of testing scenarios for the specific CAV under test. However, most existing methods often apply only to relatively simple scenarios, leaving the challenge of handling high-dimensional scenarios unsolved. The difficulty in adaptively generating high-dimensional scenarios stems from the compounded effects of the curse of rarity (CoR) and the curse of dimensionality (CoD) \cite{liu2022curse}. The CoR indicates that, due to the rarity of safety-critical events, the volume of data needed for sufficient information grows exponentially. The CoD pertains to the dimensionality of variables representing realistic scenarios, causing computational costs to escalate exponentially with the increase in scenario dimensions. Due to the CoR and CoD challenges, most existing scenario-based testing approaches are limited to short scenario segments with few background road users, involving low-dimensional decision variables that fail to capture the full complexity and variability of the real-world driving environment \cite{feng2020parti,feng2020partii,feng2020safety,zhao2016accelerated,zhao2017accelerated}. Towards addressing this challenge, our previous work introduced the naturalistic and adversarial driving environment (NADE) method capable of generating high-dimensional highway driving scenarios \cite{feng2021intelligent}. However, the NADE overlooked the performance gap between diverse CAVs and the SM, potentially impeding the testing process.

To address this problem, we develop an adaptive testing environment (AdaTE) generation method that enhances evaluation robustness by employing multiple SMs, while optimizing the combination coefficients of these SMs to improve evaluation efficiency, as shown in Fig.~\ref{fig:multiple_SMs}. In NADE, if the unsafe states of the SM can not cover all the unsafe states of the CAV under test, then the crash rate of this CAV might be underestimated. This is because NADE will rarely test crash scenarios containing unsafe states not covered by the SM. We will demonstrate such cases in Subsection \ref{subsec:testing_evaluation_results}. Using multiple SMs can broaden the coverage of unsafe states, enabling AdaTE to test various CAVs unbaisedly. In the absence of any information about the particular CAV under test, using SMs with average combination coefficients might be the most suitable approach. However, this could reduce evaluation efficiency since the resulting NADE is not tailored to any specific CAV. To improve evaluation efficiency, we optimize the combination coefficients of SMs through adaptive testing, which is formulated as a regression task using quadratic programming (QP). However, efficiently obtaining the regression target through reinforcement learning (RL) is highly challenging, as the regression target represents the prediction of crash probabilities. This is primarily due to the CoR that critical information such as crash events is rare in NDE, resulting in extremely sparse rewards. To tackle this challenge, we propose the dense reinforcement learning (DenseRL) method, extending the dense deep RL method in \cite{feng2023dense} to tabular RL. The DenseRL method, coupled with a newly designed adaptive policy, can efficiently learn the regression target. Essentially, our approach focuses on learning the values of critical state-action pairs exhibiting significant surrogate-to-real gaps. Here, the surrogate-to-real gaps represent the safety performance differences between SMs and the CAV under test. To validate our method, the high-dimensional overtaking scenarios are investigated. The results demonstrate that our approach achieves higher evaluation efficiency compared to both NDE and NADE.

The subsequent sections of this paper are structured as follows. Section \ref{sec:problem_formulation} furnishes foundational knowledge for testing CAVs in NDE and NADE, and then formulates the problem of adaptive testing in high-dimensional scenarios as a regression problem that optimizes the combination coefficients of multiple SMs. Towards addressing this problem, the AdaTE is developed in Section \ref{sec:methods}, where the DenseRL method is proposed to efficiently learn the regression target, and then the regression problem is solved using QP. The theoretical analysis for the convergence of DenseRL method is established in Section \ref{sec:theoretical_analysis}. To validate the effectiveness of the proposed method, Section \ref{sec:results} provides empirical results from testing CAVs in the high-dimensional overtaking scenarios. Finally, Section \ref{sec:conclusion} concludes the paper and discusses future research.

\section{Problem Formulation}
\label{sec:problem_formulation}

In this section, the preliminary knowledge for testing CAVs in NDE and NADE is provided in Subsection \ref{subsec:NDE_testing} and \ref{subsec:NADE_testing}, respectively. Then the adaptive testing problem will be formulated in Subsection \ref{subsec:adaptive_testing}. The list of abbreviations is shown in Table \ref{tab:abbr}. Summary of notation is listed in Table \ref{tab:notation}.

\begin{table}[!t]
  \centering
  \renewcommand{\arraystretch}{1.1}
  \caption{List of Abbreviations.}
  \label{tab:abbr}
  \begin{tabular}{cl}
    \hline
    Abbreviation & Definition \\
    \hline
    AAR & average acceleration ratio \\
    AdaTE & adaptive testing environment \\
    ASD & average sliding difference \\
    AV & automated vehicle \\
    BV & background vehicle \\
    CAV & connected and automated vehicle \\
    CoD & curse of dimensionality \\
    CoR & curse of rarity \\
    DenseRL & dense reinforcement learning \\
    FVDM & full velocity difference model \\
    IDM & intelligent driver model \\
    IF & importance function \\
    LV & leading vehicle \\
    NDE & naturalistic driving environment \\
    NADE & naturalistic and adversarial driving environment \\
    QP & quadratic programming \\
    RHW & relative half-width \\
    RL & reinforcement learning \\
    SM & surrogate model \\
    \hline 
  \end{tabular}
\end{table}

\begin{table*}[t]
  \centering
  \scriptsize
  \renewcommand{\arraystretch}{1.1}
  \caption{Summary of Notation.}
  \label{tab:notation}
  \begin{tabularx}{.9\textwidth}{cp{2.8cm}|cp{3.9cm}|cp{3.1cm}}
    \hline
    Notation             & Definition                                                              & Notation                                          & Definition                                            & Notation                           & Definition                                                               \\
    \hline
    $\bm{a},\bm{a}_t$             & action, action at time $t$   & $\mathcal{Q}$    & function space of all IFs                   & $\bm{x},\bm{X}$                                             & scenario, random variable of $\bm{x}$              \\
    $a_{\min},a_{\max}$ & min and max accelerations    & $\mathcal{Q}_J$  & function space spanned by $J$ IFs           & $v_{\text{LV}},v_{\text{BV}},v_{\text{AV}}$ & velocities of LV, BV, AV                      \\
    $\bm{A}_t$               & action random variable       & $Q,Q^*$          & state-action value function, optimal $Q$    & $x_{\text{LV}},x_{\text{BV}},x_{\text{AV}}$ & positions of LV, BV, AV                       \\
    $\mathcal{A}$       & action space                 & $Q^{(t)}$        & $Q$ at $t$-th iteration                     & $\bm{\alpha},\alpha_j$                                 & combination coefficients                      \\
    $F$                 & crash event                  & $Q_j$   & $j$-th $Q$                     & $\gamma$                                          & discount ratio                                \\
    $\mathcal{F}$       & $\sigma$-algebra             & $Q_{\bm{\alpha}}$       & $\bm{\alpha}$-combination of $Q_j$               & $\delta_t$                                        & temporal difference error                          \\
    $g$                 & surrogate-to-real gap        & $r,R$            & reward, random variable of $r$              & $\Delta$                                          & sliding stride                                \\
    $\mathbb{I}$        & indicator function           & $R_1$            & range between LV and BV                     & $\eta$                                            & adaptive policy                               \\
    $J$                 & total number of IFs          & $R_2$            & range between BV and AV                     & $\mu$                                             & crash rate in NDE                             \\
    $n$                 & total number of tests        & $\dot{R}_1$      & range rate between LV and BV                & $\hat{\mu}_n$                                     & estimation of $\mu$ in NDE                    \\
    $N(\bm{s},\bm{a})$            & visit count of $(\bm{s},\bm{a})$       & $\dot{R}_2$      & range rate between BV and AV                & $\hat{\mu}_q$                                     & estimation of $\mu$ in NADE                   \\
    $\mathbb{N}$        & set of natural numbers       & $\mathbb{R}$     & set of real numbers                         & $\nu_t$                                           & learning rate at time $t$                     \\
    $p$                 & naturalistic distribution    & $\bm{s}, \mathcal{S}$ & state, state space                          & $\sigma_{q}^2$                                    & asymptotic variance of $\hat{\mu}_{q}$        \\
    $\mathbb{P}$        & probability measure          & $\bm{s}_t,\bm{S}_t$        & state at time $t$, random variable of $s_t$ & $\sigma_{q_{\bm{\alpha}}}^2$                             & asymptotic variance of $\hat{\mu}_{q_{\bm{\alpha}}}$ \\
    $P$                 & state transition probability & $t,T$            & time step, time horizon                     & $\phi$                                            & naturalistic policy                           \\
    $q,q_j,q^*$         & IF, $j$-th IF, optimal IF    & $V,V^*$          & state value function, optimal $V$           & $\psi,\psi_j,\psi_{\bm{\alpha}}$                         & importance policies                           \\
    $q_{\bm{\alpha}}$          & mixture IF                   & $\mathcal{X}$    & scenario space                              & $\omega,\Omega$                                   & state-action pair and space                  \\
    \hline
  \end{tabularx}
\end{table*}

\subsection{Naturalistic Driving Environment Testing}
\label{subsec:NDE_testing}

Let $\bm{x}:=(\bm{s}_0,\bm{a}_0,\dots,\bm{s}_{T-1},\bm{a}_{T-1},\bm{s}_T)\in\mathcal{X}$ denote the testing scenario,
where $\bm{s}_t\in\mathcal{S}$ is the state of the CAV and background vehicles (BVs) at time $t$, $\bm{a}_t\in\mathcal{A}$ is the action of BVs at time $t$, $T$ is the time horizon, and $\mathcal{X}$ is the set of all feasible scenarios. Consider the probability space $(\mathcal{X},\mathcal{F},\mathbb{P})$, where $\mathcal{F}:=2^{\mathcal{X}}$ is the $\sigma$-algebra, $\mathbb{P}(\{\bm{x}\}):= p(\bm{x}),~\forall\bm{x}\in\mathcal{X}$ is the probability measure, and $p$ is the naturalistic distribution. Denote the crash event between the CAV and BVs as $F:=\{\bm{x}\in\mathcal{X}:\bm{s}_T\in\mathcal{S}_{\mathrm{crash}}\}\in\mathcal{F}$, where $\mathcal{S}_{\mathrm{crash}}$ is the set of crash states. Then the crash rate in NDE is given by $\mu:=\mathbb{P}(F)=\mathbb{E}_p[\mathbb{I}_F(\bm{X})]$, where $\mathbb{I}_F$ is the indicator function of $F$, and $\bm{X}:\bm{x}\mapsto\bm{x}$, $\forall\bm{x}\in\mathcal{X}$ is the scenario random variable. According to Monte Carlo theory \cite{owen2013monte}, the crash rate can be estimated in NDE as
\begin{equation}
  \label{eq:CR_MC_NDE}
  \hat{\mu}_n:=\frac{1}{n}\sum_{i=1}^n{\mathbb{I}_F(\bm{X}_i)},~ \bm{X}_i\sim p,
\end{equation}
where $n$ is the number of tests, and $\bm{X}_i$ are scenario random variables sampled i.i.d. from $p$.

\subsection{Naturalistic and Adversarial Driving Environment Testing}
\label{subsec:NADE_testing}

The evaluation efficiency of NDE suffers severely from the CoR \cite{liu2022curse}. Using importance sampling method to replace the naturalistic distribution with the importance function (IF) is helpful to improve the evaluation efficiency \cite{feng2020parti,feng2020partii,zhao2016accelerated,zhao2017accelerated}. However, the importance sampling method can not be directly applied in high-dimensional scenarios because of the CoD \cite{au2003important}. Therefore, the NADE method has been proposed to only control critical variables at critical moments, while keeping other variables with their naturalistic distributions \cite{feng2021intelligent}. Leveraging such importance function $q$, the crash rate can be estimated in NADE as
\begin{equation}
  \label{eq:CR_NADE}
  \hat{\mu}_q
  :=\frac{1}{n}\sum_{i=1}^n\frac{\mathbb{I}_F(\bm{X}_i)p(\bm{X}_i)}{q(\bm{X}_i)},~\bm{X}_i\sim q.
\end{equation}

\subsection{Adaptive Testing}
\label{subsec:adaptive_testing}

Although the NADE has shown great potential for testing CAVs efficiently, one crucial issue arises when testing diverse CAVs. The performance of NADE strongly relies on the selected importance function, which may not be suitable for various CAVs, leading to catastrophic failures. Towards solving this issue, the goal of adaptive testing is to improve the robustness of NADE for diverse CAVs, while keeping the evaluation efficiency. One approach is to solve the optimization problem that minimizes the estimation variance of NADE over the function space $\mathcal{Q}$ that incorporates all possible importance functions, i.e.,
\begin{equation}
  \label{eq:optim_adaptive_scenario_generation}
  \min_{q\in\mathcal{Q}}~\sigma_q^2:=\mathrm{Var}_q\left(\frac{\mathbb{I}_F(\bm{X})p(\bm{X})}{q(\bm{X})}\right).
\end{equation}
By optimizing $q$ in $\mathcal{Q}$, the importance function could be customized for the specific CAV under test, thus improving the evaluation efficiency for diverse CAVs.

Solving the optimization problem (\ref{eq:optim_adaptive_scenario_generation}) is highly challenging, because the optimization space $\mathcal{Q}$ is a general function space. To address this issue, we propose to reduce the optimization space $\mathcal{Q}$ to the function space spanned by multiple importance functions, which can be formulated as $\mathcal{Q}_J:=\{q_{\bm{\alpha}}\in\mathcal{Q}:\textbf{1}^\top\bm{\alpha}=1,~\bm{\alpha}\geqslant\textbf{0}\}\subset\mathcal{Q}$,
where $q_{\bm{\alpha}}$ is the mixture importance function with mixture importance policy $\psi_{\bm{\alpha}}:=\sum_{j=1}^J \alpha_j\psi_j$, $\psi_j:\mathcal{S}\times\mathcal{A}\to\mathbb{R}_{\geqslant0}$ are importance policies, $\alpha_j$ are combination coefficients, $\bm{\alpha}:=[\alpha_1,\dots,\alpha_J]^\top$, and $J$ is the number of importance functions. We note that these importance functions could be obtained from multiple SMs \cite{feng2021intelligent}.
With $\mathcal{Q}_J$ in place of $\mathcal{Q}$, the optimization problem (\ref{eq:optim_adaptive_scenario_generation}) can be simplified as
\begin{equation}
  \label{eq:optim_problem_spanned}
  \min_{q_{\bm{\alpha}}\in\mathcal{Q}_J}~\sigma_{q_{\bm{\alpha}}}^2:=\mathrm{Var}_{q_{\bm{\alpha}}}\left(\frac{\mathbb{I}_F(\bm{X})p(\bm{X})}{q_{\bm{\alpha}}(\bm{X})}\right).
\end{equation}

Then our goal is to optimize $q_{\bm{\alpha}}$ towards the optimal importance function $q^*$, which can be approximated by optimizing $\psi_{\bm{\alpha}}$ towards the optimal importance policy $\psi^*$. According to importance sampling theory \cite{owen2013monte}, the optimal importance policy is given by $\psi^*(\bm{a}|\bm{s})=Q^*(\bm{s},\bm{a})\phi(\bm{a}|\bm{s})/V^*(\bm{s})$, where $Q^*(\bm{s},\bm{a}):=\mathbb{P}(F|\bm{S}=\bm{s},\bm{A}=\bm{a})$ is the maneuver challenge that represents the crash probability given current state-action pair $(\bm{s},\bm{a})$, $\bm{S}:\bm{x}\mapsto\bm{s}$, $\forall\bm{x}\in\mathcal{X}$ is the state random variable, $\bm{A}:\bm{x}\mapsto\bm{a}$, $\forall\bm{x}\in\mathcal{X}$ is the action random variable, $\phi(\bm{a}|\bm{s})$ is the naturalistic policy of BVs' action $\bm{a}$ given current state $\bm{s}$, and $V^*(\bm{s}):=\sum_{\bm{a}\in\mathcal{A}}Q^*(\bm{s},\bm{a})\phi(\bm{a}|\bm{s})$ is the criticality. Then the optimization problem (\ref{eq:optim_problem_spanned}) can be simplified as a regression task via QP, i.e.,
\begin{equation}
  \label{eq:simplified_optim_problem}
  \begin{aligned}
    \min_{\bm{\alpha}\in\mathbb{R}^J}~&\frac{1}{2}\sum_{\bm{s}\in\mathcal{S},\bm{a}\in\mathcal{A}}\big[Q^*(\bm{s},\bm{a})-Q_{\bm{\alpha}}(\bm{s},\bm{a})\big]^2\\
    \text{s.t.}~~&\textbf{1}^\top\bm{\alpha}=1,~\bm{\alpha}\geqslant\textbf{0},
  \end{aligned}
\end{equation}
where $Q_{\bm{\alpha}}:=\sum_{j=1}^{J}\alpha_jQ_j$ is the mixture maneuver challenge, and $Q_j$ are surrogate maneuver challenges associated with importance policies $\psi_j$. The key for solving this optimization problem lies in efficiently obtaining $Q^*$, which can be formulated as a RL problem (see Subsection \ref{subsec:dense_RL}). However, learning $Q^*$ through RL in NDE is highly challenging due to the CoR that the critical information such as crash events is extremely rare. Moreover, since our goal is to optimize the combination coefficients, accurately computing $Q^*$ across the entire state-action space is unnecessary, as it requires large number of tests and thereby compromises optimization efficiency. We will address these challenges in the forthcoming Section \ref{sec:methods}.

\section{Methods}
\label{sec:methods}

To address the CoR, we first propose in Subsection \ref{subsec:dense_RL} the dense reinforcement learning method. To facilitate DenseRL method for adaptive testing, a new adaptive policy with high sample efficiency is designed in Subsection \ref{subsec:adaptive_policy}. Then the regression problem that optimizes combination coefficients will be solved via QP in Subsection \ref{subsec:regression}. Finally, Subsection \ref{subsec:AdaTE_generation_alg} summarizes the AdaTE generation algorithm.

\subsection{Dense Reinforcement Learning}
\label{subsec:dense_RL}

\begin{figure}[!t]
  \centering
  \includegraphics[width=7.4cm]{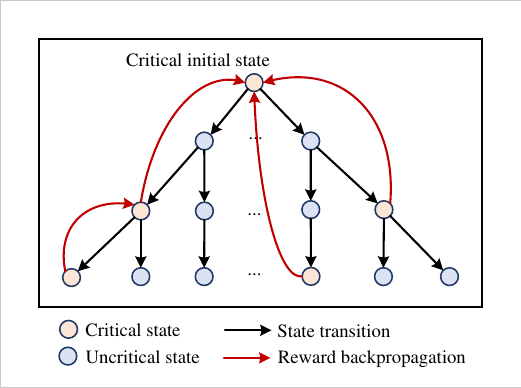}
  \caption{Illustration of the dense reinforcement learning method.}
  \label{fig:DenseRL}
\end{figure}

The problem to find $Q^*$ can be formulated as a RL problem. Define $\mathcal{M}:=(\mathcal{S},\mathcal{A},R,P,\gamma)$ as the Markov decision process, where $P$ is the state transition probability, $R$ is the reward function, $R(\bm{s}):=\mathbb{I}_{\mathcal{S}_{\mathrm{crash}}}(\bm{s})$, $\forall\bm{s}\in\mathcal{S}$, and $\gamma\in(0, 1)$ is the discount factor. Then $Q^*$ can be represented as the state-action value function given by
\begin{equation}
  Q^*(\bm{s},\bm{a})=\mathbb{E}_p\left[\sum_{\tau=t+1}^T\gamma^{\tau-t-1} R_\tau\Bigg|\bm{S}_t=\bm{s},\bm{A}_t=\bm{a}\right],
\end{equation}
where $t$ is the time step of $(\bm{s},\bm{a})$, $R_\tau:=R$.
However, learning $Q^*$ by RL in NDE faces the CoR, because the informative data (i.e., critical states and actions) in NDE is rare and the rewards (i.e., crash events) are extremely sparse. To address this challenge, we propose the dense reinforcement learning method, following the similar ideas in \cite{feng2023dense}. As shown in Fig.~\ref{fig:DenseRL}, the core concept of DenseRL is to start with critical initial states, use off-policy learning mechanism, edit the Markov chains by removing the uncritical states and reconnecting the critical states, and then backpropagate the reward along the edited Markov chains.

Initially, we set $Q(\bm{s},\bm{a})=0$, $\forall\bm{s}\in\mathcal{S}$, $\bm{a}\in\mathcal{A}$. To optimize $Q$ towards $Q^*$, DenseRL tries to minimize the Bellman error $\bar{\delta}:=\mathcal{B}_\phi Q-Q$, where $\mathcal{B}_\phi$ is the Bellman backup operator with $\mathcal{B}_\phi Q(\bm{s},\bm{a}):=\mathbb{E}_\phi[R_{t+1}+\gamma Q(\bm{S}_{t+1},\bm{A}_{t+1})|\bm{S}_t=\bm{s},\bm{A}_t=\bm{a}]$ \cite{sutton2018reinforcement,li2023reinforcement}. Let $\mathcal{S}_c:=\{\bm{s}\in\mathcal{S}:\bar{V}(\bm{s})>0\}$ denote the set of critical states, where $\bar{V}(\bm{s}):=\sum_{\bm{a}\in\mathcal{A}}\bar{Q}(\bm{s},\bm{a})\phi(\bm{a}|\bm{s})$ and $\bar{Q}:=(1/J)\sum_{j=1}^JQ_j$. Then for each training iteration, the initial state will be sampled uniformly from $\mathcal{S}_c$, thereafter following an appropriate behavior policy (e.g., the uniform policy). For each transition $(\bm{S}_t,\bm{A}_t,R_{t+1},\bm{S}_{t+1})$, DenseRL learns $Q$ by the following update rule:
\begin{equation}
  \label{eq:update_rule}
  Q(\bm{S}_t,\bm{A}_t)\leftarrow Q(\bm{S}_t,\bm{A}_t)+\nu_t\delta_t\mathbb{I}_{\mathcal{S}_c}(\bm{S}_t),
\end{equation}
where $\nu_t$ is the learning rate, $\delta_t:=\hat{\mathcal{B}}_\phi Q(\bm{S}_t,\bm{A}_t)-Q(\bm{S}_t,\bm{A}_t)$ is the temporal difference error, and $\hat{\mathcal{B}}_\phi$ is the Bellman evaluation operator with $\hat{\mathcal{B}}_\phi Q(\bm{S}_t,\bm{A}_t):=R_{t+1}+\gamma\mathbb{E}_\phi[Q(\bm{S}_{t+1},\bm{A}_{t+1})|\bm{S}_{t+1}]$.


\subsection{Adaptive Policy Design}
\label{subsec:adaptive_policy}

In adaptive testing, our focus is primarily on state-action pairs that exhibit significant surrogate-to-real gaps. This emphasis is not effectively utilized when employing DenseRL with a uniform policy, which can result in diminished learning efficiency. Here, the surrogate-to-real gap aims to measure the gap between $Q_{\bm{\alpha}}$ and $Q$, which is defined for all $\bm{s}\in\mathcal{S}$, $\bm{a}\in\mathcal{A}$ as
\begin{equation}
  \label{eq:surrogate_to_real_gap}
  g(Q||Q_{\bm{\alpha}}):=
  \begin{cases}
    \dfrac{|Q-Q_{\bm{\alpha}}|}{Q_{\bm{\alpha}}},&\mathrm{if}~Q_{\bm{\alpha}}>0,\\
    0,&\mathrm{if}~Q=Q_{\bm{\alpha}}=0,\\
    +\infty,&\mathrm{if}~Q>Q_{\bm{\alpha}}=0.\\
  \end{cases}
\end{equation}
To increase the learning efficiency, we devise a new adaptive policy based on the surrogate-to-real gap and the probabilistic upper confidence tree bound \cite{silver2016mastering}. Specifically, the adaptive policy is defined as
\begin{equation}
  \label{eq:adaptive_policy}
  \eta(\bm{a}|\bm{s}):=
  \begin{cases}
    1,&\mathrm{if}~\bm{a}=\argmax\limits_{\bm{a}'\in\mathcal{A}} U(\bm{s},\bm{a}'),\\
    0,&\mathrm{otherwise},
  \end{cases}
\end{equation}
where $U(\bm{s},\bm{a}):=U'(\bm{s},\bm{a})\phi(\bm{a}|\bm{s})$, and
\begin{equation}
  U'(\bm{s},\bm{a}):=g(Q||Q_{\bm{\alpha}})(\bm{s},\bm{a})+c\frac{\sqrt{\sum_{\bm{a}'\in\mathcal{A}}N(\bm{s},\bm{a}')}}{1+N(\bm{s},\bm{a})}.
\end{equation}
Here, $c$ is a constant that determines the degree of exploration, and $N(\bm{s},\bm{a})$ is the visit count of $(\bm{s},\bm{a})$, $\forall\bm{s}\in\mathcal{S}$, $\bm{a}\in\mathcal{A}$.

\subsection{Combination Coefficient Optimization}
\label{subsec:regression}

Let $\mathcal{D}$ denote the set of visited critical state-action pairs, then the combination coefficients can be optimized by solving the following regression problem:
\begin{equation}
  \label{eq:regression_QP}
  \begin{aligned}
    \min_{\bm{\alpha}\in\mathbb{R}^J}&~\frac{1}{2}\sum_{(\bm{s},\bm{a})\in\mathcal{D}}\big[Q(\bm{s},\bm{a})-Q_{\bm{\alpha}}(\bm{s},\bm{a})\big]^2\\
    \text{s.t.}~&\textbf{1}^\top\bm{\alpha}=1,~\bm{\alpha}\geqslant\textbf{0},
  \end{aligned}
\end{equation}
where $Q$ is learned by DenseRL with the adaptive policy. This regression problem (\ref{eq:regression_QP}) is a QP, which can be solved by standard convex optimization tools such as CVXOPT \cite{andersen2004cvxopt}.

\subsection{Adaptive Testing Environment Generation Algorithm}
\label{subsec:AdaTE_generation_alg}

\begin{algorithm}[!t]
  \label{alg:AdaTE}
  \caption{Adaptive testing environment generation via dense reinforcement learning}
  \KwIn{naturalistic distribution $p$, surrogate maneuver challenges $Q_j$, $j=1,\dots,J$, max simulation time $T$}
  \KwOut{combination coefficients $\bm{\alpha}$}
  Initialize $Q(\bm{s},\bm{a})=0$, $N(\bm{s},\bm{a})=0$, $\forall\bm{s}\in\mathcal{S}$, $\bm{a}\in\mathcal{A}$\;
  Initialize $i=0$, $\Delta=10$, $\bm{\alpha}=\textbf{1}/J$\;
  Initialize $c=2$, \textit{termination} = False\;
  \While{not termination}{
    Set $i\leftarrow i+1$\;
    Sample initial state $\bm{s}$ uniformly from $\mathcal{S}_c$\;
    Set $r\leftarrow 0$, $t\leftarrow 0$\;
    \While{$r=0$ and $t<T$}{
      Set $t\leftarrow t+1$\;
      Sample $\bm{a}$ from the adaptive policy $\eta$ given by Eq.~(\ref{eq:adaptive_policy})\;
      Set $N(\bm{s},\bm{a})\leftarrow N(\bm{s},\bm{a})+1$\;
      Take action $\bm{a}$, and observe $\bm{s}',r$\;
      Update $Q(\bm{s},\bm{a})$ according to Eq.~(\ref{eq:update_rule})\;
      Set $\bm{s}\leftarrow\bm{s}'$\;
    }
    Update $\bm{\alpha}$ by solving the QP in Eq.~(\ref{eq:regression_QP}) (e.g., via CVXOPT \cite{andersen2004cvxopt})\;
    Update \textit{termination} according to Eq.~(\ref{eq:ASD})\;
  }
  Return $\bm{\alpha}$\;
\end{algorithm}

By utilizing DenseRL, the combination coefficients can be optimized for the particular CAV under test, resulting in the generation of AdaTE. This process is outlined in Algorithm \ref{alg:AdaTE}. The termination criterion is when the average sliding difference (ASD) falls below a predetermined threshold (e.g., 0.02), which is defined as
\begin{equation}
  \label{eq:ASD}
  \mathrm{ASD}(k):=\frac{1}{J}\sum_{j=1}^{J}\left|\sum_{k'=k-\Delta+1}^{k}\left[\alpha_j^{(k')}-\alpha_j^{(k'-\Delta)}\right]\right|,
\end{equation}
where $k\in\mathbb{N}_{>0}$ is the number of tests, $\Delta\in\mathbb{N}_{>0}$ is the sliding stride (e.g., $\Delta=10$), $\alpha_j^{(k)}$ are combination coefficients of the $k$-th iteration, and we set $\alpha_j^{(k)}:=\alpha_j^{(1)}$ for $k<1$.

\section{Theoretical Analysis}
\label{sec:theoretical_analysis}

This section will provide theoretical analysis of the proposed DenseRL method. Specifically, we prove the convergence of DenseRL, i.e., $Q^{(t)}$ converges to $Q^*$ with probability one, where $Q^{(t)}$ represents the $Q$ function at $t$-th iteration and $t\in\mathbb{N}_{\geqslant0}$. The proof is based on the following lemma \cite{jaakkola1993convergence}.
\begin{lemma}
  \label{lem:stochastic_approximation}
  Consider the stochastic process $(\nu_t,\Delta_t,F_t)$, $t\in\mathbb{N}_{\geqslant0}$, where $\nu_t$, $\Delta_t$, $F_t:\Omega\to\mathbb{R}$ satisfy $\Delta_{t+1}(\omega)=[1-\nu_t(\omega)]\Delta_t(\omega)+\nu_t(\omega)F_t(\omega)$, $\omega\in\Omega$. Let $\mathcal{F}_t$ be a sequence of increasing $\sigma$-fields such that $\nu_0$ and $\Delta_0$ are $\mathcal{F}_0$-measurable and $\nu_t$, $\Delta_t$ and $F_{t-1}$ are $\mathcal{F}_t$-measurable, $t\in\mathbb{N}_{>0}$. Then $\Delta_t$ converges to zero with probability one under the following Assumption \ref{asm:finite_space}, \ref{asm:learning_rate}, \ref{asm:bounded_expectation} and \ref{asm:bounded_variance}.
  \begin{assumption}
    \label{asm:finite_space}
    The set $\Omega$ is finite.
  \end{assumption}
  \begin{assumption}
    \label{asm:learning_rate}
    $\nu_t(\omega)\in[0,1]$, $t\in\mathbb{N}_{\geqslant0}$, $\sum_t\nu_t(\omega)=\infty$, $\sum_t\nu_t^2(\omega)<\infty$, $\forall\omega\in\Omega$.
  \end{assumption}
  \begin{assumption}
    \label{asm:bounded_expectation}
    $\|\mathbb{E}[F_t|\mathcal{F}_t]\|_\infty\leqslant\gamma\|\Delta_t\|_\infty$, where $\gamma\in(0,1)$, $t\in\mathbb{N}_{\geqslant0}$.
  \end{assumption}
  \begin{assumption}
    \label{asm:bounded_variance}
    $\mathrm{Var}(F_t(\omega)|\mathcal{F}_t)\leqslant C(1+\|\Delta_t\|_\infty)^2$, $C>0$, $t\in\mathbb{N}_{\geqslant0}$.
  \end{assumption}
\end{lemma}

\begin{IEEEproof}
  See \cite{jaakkola1993convergence}.
\end{IEEEproof}

Leveraging Lemma \ref{lem:stochastic_approximation}, we are now in position to prove the following theorem.

\begin{theorem}
  \label{thm:convergence}
  The DenseRL algorithm given by Subsection \ref{subsec:dense_RL} converges with probability one to $Q^*$ under the following Assumption \ref{asm:finite_state_action}, \ref{asm:learning_rate_dense}, \ref{asm:initial_zero} and \ref{asm:coverage}.
  \begin{assumption}
    \label{asm:finite_state_action}
    The sets $\mathcal{S}$ and $\mathcal{A}$ are finite.
  \end{assumption}
  \begin{assumption}
    \label{asm:learning_rate_dense}
    $\nu_t(\bm{s},\bm{a})\in[0,1]$, $t\in\mathbb{N}_{\geqslant0}$, $\sum_t\nu_t(\bm{s},\bm{a})=\infty$, $\sum_t\nu_t^2(\bm{s},\bm{a})<\infty$, $\forall\bm{s}\in\mathcal{S}_c$, $\bm{a}\in\mathcal{A}$.
  \end{assumption}
  \begin{assumption}
    \label{asm:initial_zero}
    $Q^{(0)}(\bm{s},\bm{a})=0$, $\forall\bm{s}\in\mathcal{S}$, $\bm{a}\in\mathcal{A}$.
  \end{assumption}
  \begin{assumption}
    \label{asm:coverage}
    $\bar{V}(\bm{s})>0$ whenever $V^*(\bm{s})>0$.
  \end{assumption}
\end{theorem}

\begin{IEEEproof}
  The correspondence to Lemma \ref{lem:stochastic_approximation} follows from associating $\Omega$ with the state-action space $\mathcal{S}\times\mathcal{A}$, $\omega$ with the state-action pair $(\bm{s},\bm{a})$, $\nu_t(\omega)$ with the learning rate $\nu_t(\bm{s},\bm{a})$, $\Delta_t(\omega)$ with $Q^{(t)}(\bm{s},\bm{a})-Q^*(\bm{s},\bm{a})$, and $\mathcal{F}_t$ with the $\sigma$-field generated by random variables $\{Q^{(0)},\bm{S}_0,\bm{A}_0,\nu_0,R_1,\dots,\bm{S}_t,\bm{A}_t,\nu_t\}$. Then Theorem \ref{thm:convergence} can be proved by verifying Assumption \ref{asm:finite_space}, \ref{asm:learning_rate}, \ref{asm:bounded_expectation} and \ref{asm:bounded_variance} in Lemma \ref{lem:stochastic_approximation} accordingly.
  \begin{enumerate}
    \item Verify Assumption \ref{asm:finite_space}. Assumption \ref{asm:finite_state_action} clearly confirms that $\Omega=\mathcal{S}\times\mathcal{A}$ is finite.
    \item Verify Assumption \ref{asm:learning_rate}. Assumption \ref{asm:learning_rate} in Lemma \ref{lem:stochastic_approximation} requires that all state-action pairs be visited infinitely often \cite{melo2001convergence}. According to Assumption \ref{asm:initial_zero} and \ref{asm:coverage}, we know that $Q^{(t)}(\bm{s},\bm{a})=Q^*(\bm{s},\bm{a})=0$, $\forall\bm{s}\in\mathcal{S}_{-c}$, $\bm{a}\in\mathcal{A}$. In other words, the state-action values for uncritical states are already optimal values, and hence do not need to be visited. It is sufficient that all critical state-action pairs will be visited infinitely often, therefore the Assumption \ref{asm:learning_rate} can be verified by Assumption \ref{asm:learning_rate_dense}.
    \item Verify Assumption \ref{asm:bounded_expectation}. Rewriting Eq.~(\ref{eq:update_rule}) we get
    \begin{equation}
      Q^{(t+1)}(\omega_t)=\big[1-\nu_t(\omega_t)\big]Q^{(t)}(\omega_t)+\nu_t(\omega_t)\hat{\mathcal{B}}_\phi Q^{(t)}(\omega_t).
    \end{equation}
    Subtracting from both sides the quantity $Q^*(\omega_t)$ yields
    \begin{equation}
      \Delta_{t+1}(\omega_t)
      =\big[1-\nu_t(\omega_t)\big]\Delta_t(\omega_t)+\nu_t(\omega_t)F_t(\omega_t),
    \end{equation}
    where $F_t:=\hat{\mathcal{B}}_\phi Q^{(t)}-Q^*$. Since $\mathcal{B}_\phi$ is a $\gamma$-contraction mapping \cite{szepesvari2022algorithms}, we have
    \begin{equation}
      \begin{aligned}
        \|\mathbb{E}[F_t|\mathcal{F}_t]\|_\infty
        &=\|\mathcal{B}_\phi Q^{(t)}-Q^*\|_\infty\\
        &=\|\mathcal{B}_\phi Q^{(t)}-\mathcal{B}_\phi Q^*\|_\infty\\
        &\leqslant\gamma\|Q^{(t)}-Q^*\|_\infty
        =\gamma\|\Delta_t\|_\infty.
      \end{aligned}
    \end{equation}
    \item Verify Assumption \ref{asm:bounded_variance}. Due to the fact that the reward function is bounded, we have
    \begin{equation}
      \begin{aligned}
        \mathrm{Var}\big(F_t(\omega_t)|\mathcal{F}_t\big)
        &=\mathrm{Var}\big(\hat{\mathcal{B}}_\phi Q^{(t)}(\omega_t)-Q^*(\omega_t)|\mathcal{F}_t\big)\\
        &=\mathrm{Var}\big(\hat{\mathcal{B}}_\phi Q^{(t)}(\omega_t)|\mathcal{F}_t\big)\\
        &\leqslant C(1+\|\Delta_t\|_\infty)^2,
      \end{aligned}
    \end{equation}
    for some constant $C>0$.
  \end{enumerate}
  To verify the measurability requirements in Lemma \ref{lem:stochastic_approximation}, we note that $Q^{(t)}$ are $\mathcal{F}_t$-measurable, and thus both $\Delta_t$ and $F_{t-1}$ are $\mathcal{F}_t$-measurable. Therefore, by Lemma \ref{lem:stochastic_approximation} we know that $\Delta_t$ converges to zero with probability one, i.e., $Q^{(t)}$ converges to $Q^*$ with probability one.
\end{IEEEproof}

\begin{remark}
  The Assumption \ref{asm:finite_state_action} can be satisfied if both the state space and the action space are discretized. Similar with Assumption \ref{asm:learning_rate}, the Assumption \ref{asm:learning_rate_dense} requires that all critical state-action pairs be visited infinitely often. The Assumption \ref{asm:initial_zero} is an initialization requirement that ensures all uncritical state-action values are optimal values (i.e., 0). Moreover, the Assumption \ref{asm:coverage} means that the critical states identified by all surrogate criticalities can cover the critical states of the CAV under test, since otherwise we would omit critical states that should be explored and learned, leading to misconvergence issues.
\end{remark}

\begin{remark}
  In adaptive testing, to fulfill the requirement of Assumption \ref{asm:coverage}, the selected SMs should exhibit sufficient diversity to encompass the critical states of various CAVs.
\end{remark}

\section{Results}
\label{sec:results}

In this section, the high-dimensional overtaking scenarios will be elaborated in Subsection \ref{subsec:overtaking}. Then in Subsection \ref{subsec:testing_evaluation_results}, the testing and evaluation results in NDE, NADE and AdaTE will be presented and analyzed.

\subsection{Overtaking Scenarios}
\label{subsec:overtaking}

\begin{figure}[!t]
  \centering
  \subfigure[Four phases of overtaking scenarios.]{\includegraphics[width=8.75cm]{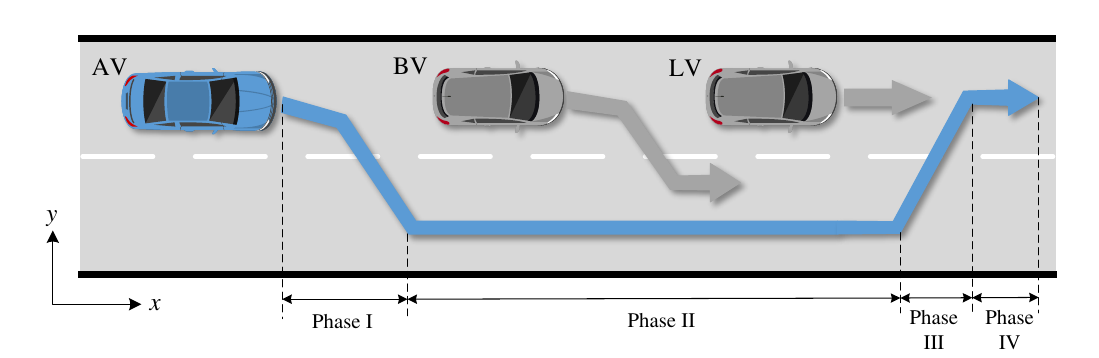}
  \label{subfig:overall_overtaking}}
  \subfigure[Passing phase of overtaking scenarios (focus of this paper).]{\includegraphics[width=8.75cm]{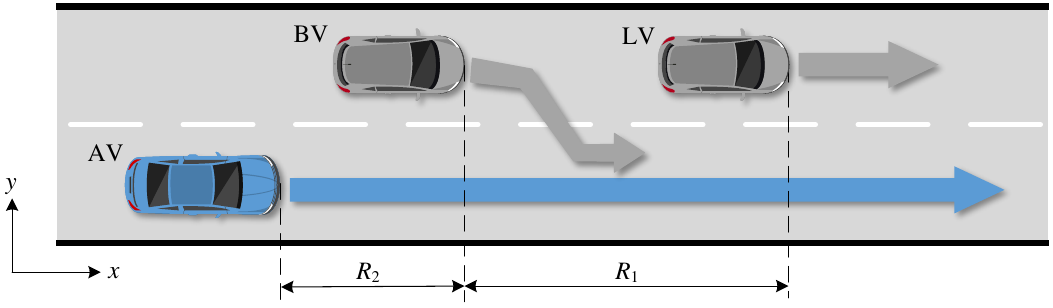}
  \label{subfig:passing_phase}}
  \caption{Illustrations of the four phases of overtaking scenarios (a) and the passing phase (Phase II) of overtaking scenarios (b). In overtaking scenarios, the AV will overtake BV and LV. In the passing phase, the AV will pass BV and LV. While AV is passing, BV may overtake LV.}
  \label{fig:overtaking}
\end{figure}

As shown in Fig.~\ref{fig:overtaking}, we study the passing phase of the high-dimensional overtaking scenarios, where a relatively slow-moving leading vehicle (LV) travels in front of the BV, while the automated vehicle (AV) is going to overtake BV and LV. Meanwhile, BV can also overtake LV, then AV may rear-end with BV, resulting in a crash. Denote the longitudinal positions and velocities of LV, BV and AV as $x_{\text{LV}}$, $x_{\text{BV}}$, $x_{\text{AV}}$, $v_{\text{LV}}$, $v_{\text{BV}}$, $v_{\text{AV}}$, respectively, then the state of the overtaking scenarios can be formulated as $\bm{s}:=[v_{\text{BV}}, R_1, \dot{R}_1, R_2, \dot{R}_2]^\top$,
where $R_1:= x_{\text{LV}}-x_{\text{BV}}$, $\dot{R}_1:= v_{\text{LV}}-v_{\text{BV}}$, $R_2:= x_{\text{BV}}-x_{\text{AV}}$, and $\dot{R}_2:= v_{\text{BV}}-v_{\text{AV}}$. The action of the overtaking scenarios is defined as the collection of accelerations of LV and BV, i.e., $\bm{a}:=[a_{\text{LV}},a_{\text{BV}}]^\top$. The maximum simulation time and time resolution are set to 20 s and 0.1 s, respectively. Typically, overtaking scenarios will exceed 1400 dimensions (201 time steps, each with 5 state variables and 2 action variables), presenting the high-dimensionality challenge.


\subsection{Testing and Evaluation Results}
\label{subsec:testing_evaluation_results}

In this subsection, we present and analyze the testing and evaluation results in NDE, NADE and AdaTE. For the generation of NDE and NADE, we use the same way as in \cite{feng2021intelligent}\footnote{Link to source code: https://github.com/michigan-traffic-lab/Naturalistic-and-Adversarial-Driving-Environment.}. To investigate the generalizability of the proposed method, we test three diverse AVs: (1) the intelligent driver model (IDM) \cite{ro2017formal}, denoted as AV-I; (2) the IDM calibrated in \cite{sangster2013application}, denoted as AV-II; (3) the RL agent trained by proximal policy optimization \cite{schulman2017proximal}, denoted as AV-III. We use three representative SMs involving normal, aggressive and conservative driving styles: (1) IDM, denoted as SM-I (same as AV-I); (2) the full velocity difference model (FVDM) \cite{ro2017formal} with $a_{\min}=-1$ m/s\textsuperscript{2}, denoted as SM-II; (3) FVDM with $a_{\min}=-6$ m/s\textsuperscript{2}, denoted as SM-III.

\begin{figure}[!t]
  \centering
  \setlength{\subfiglen}{5cm}
  \subfigure[AV-I]{\includegraphics[width=\subfiglen]{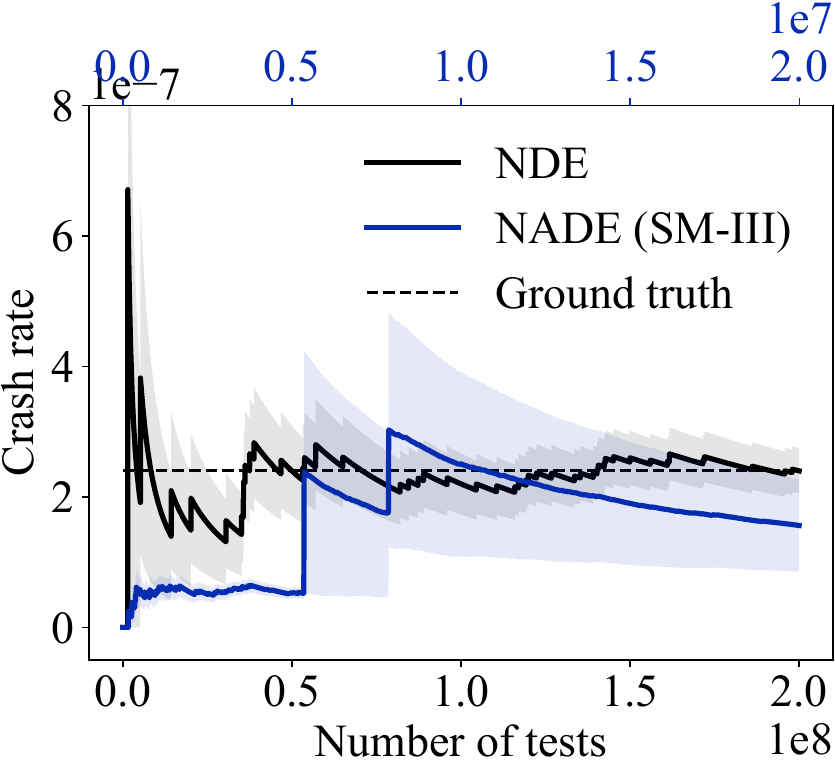}
  \label{subfig:CR_NDE_NADE_AV_1_SM_3}}
  \\
  \subfigure[AV-I]{\includegraphics[width=\subfiglen]{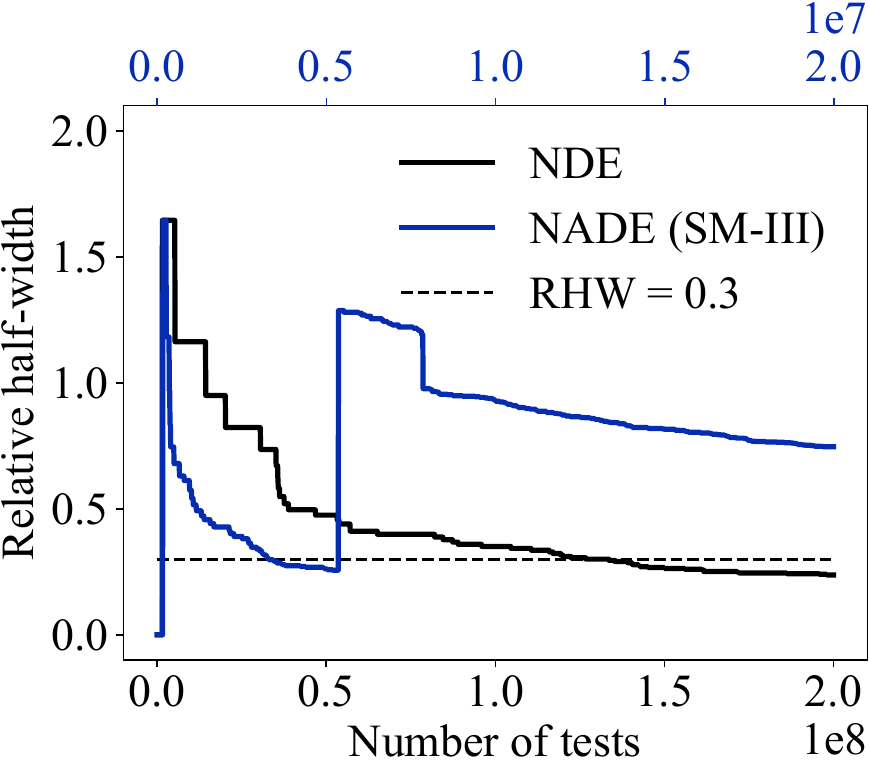}
  \label{subfig:RHW_NDE_NADE_AV_1_SM_3}}
  \caption{(a) The crash rate estimations of AV-I in NDE and the NADE where the importance function is constructed from SM-III. (b) The RHW of crash rate estimations.}
  \label{fig:CR_RHW_NDE_NADE_AV_1_SM_3}
\end{figure}



\begin{figure*}[!t]
  \centering
  \setlength{\subfiglen}{5cm}
  \subfigure[AV-I]{\includegraphics[width=\subfiglen]{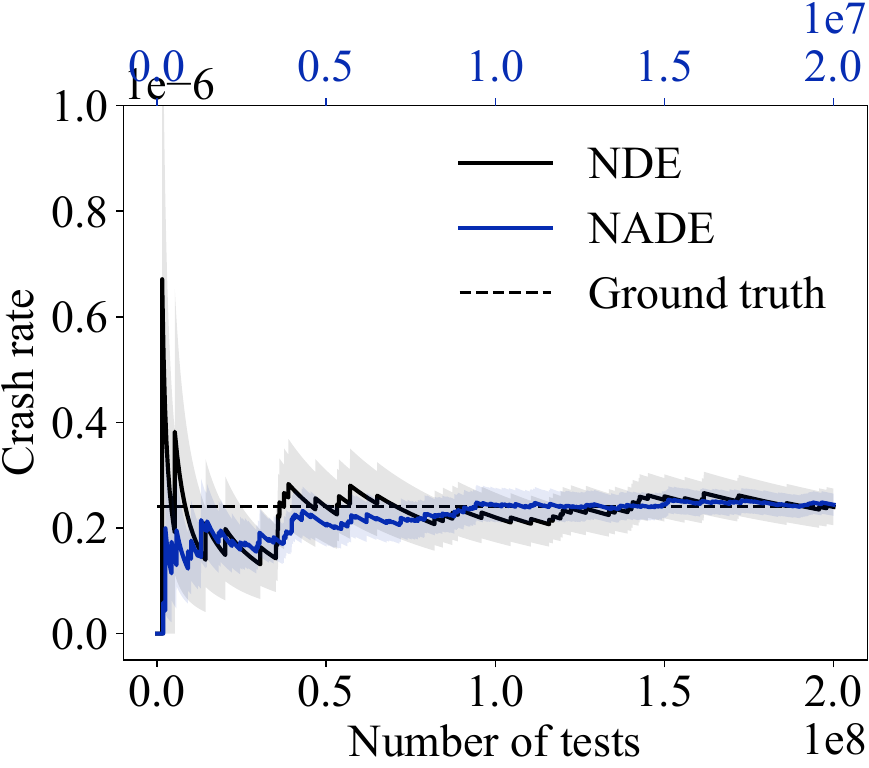}\label{subfig:CR_NDE_NADE_AV_1}}
  \quad
  \subfigure[AV-II]{\includegraphics[width=\subfiglen]{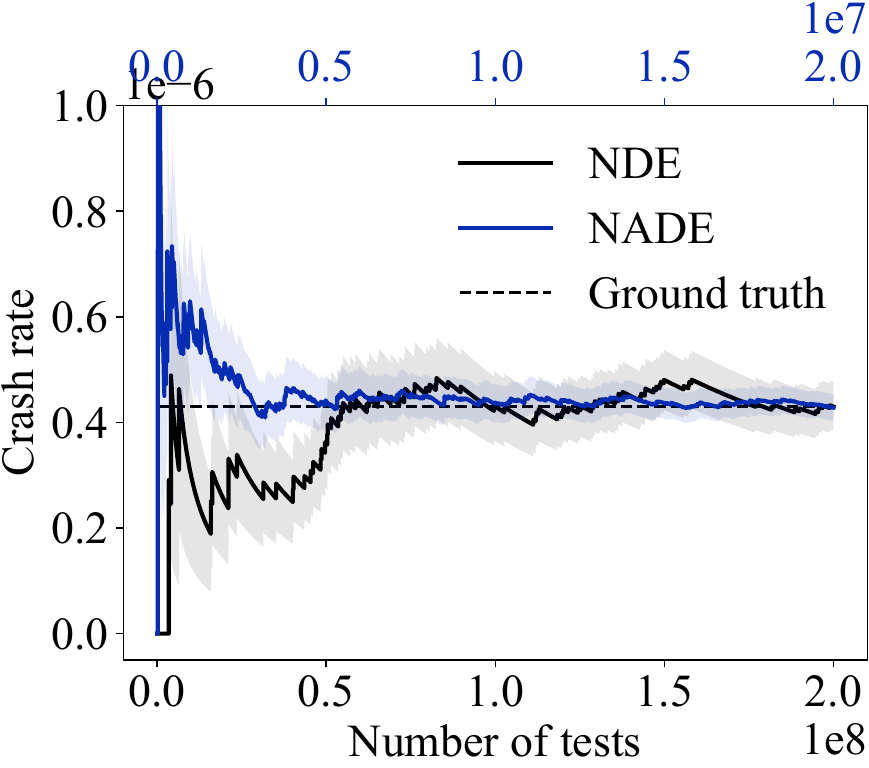}\label{subfig:CR_NDE_NADE_AV_2}}
  \quad
  \subfigure[AV-III]{\includegraphics[width=\subfiglen]{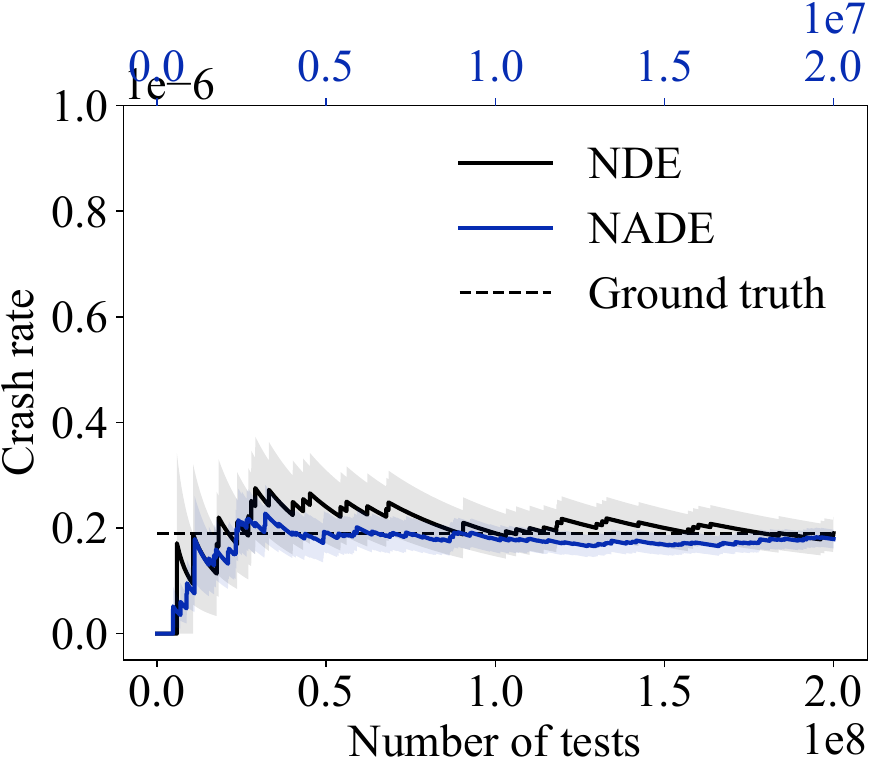}\label{subfig:CR_NDE_NADE_AV_3}}
  \\
  \subfigure[AV-I]{\includegraphics[width=\subfiglen]{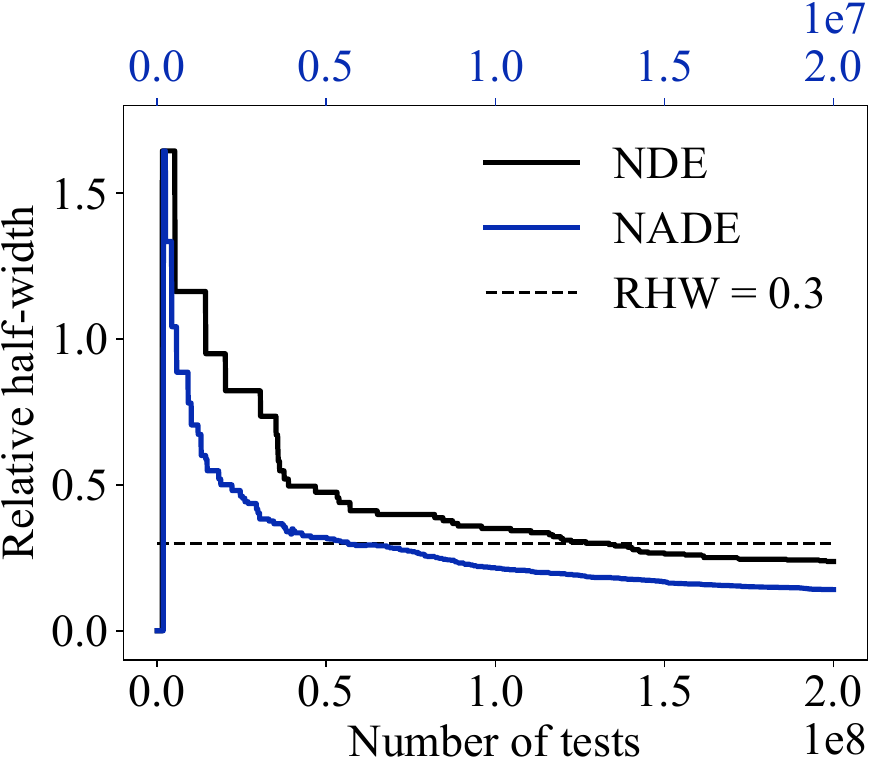}\label{subfig:RHW_NDE_NADE_AV_1}}
  \quad
  \subfigure[AV-II]{\includegraphics[width=\subfiglen]{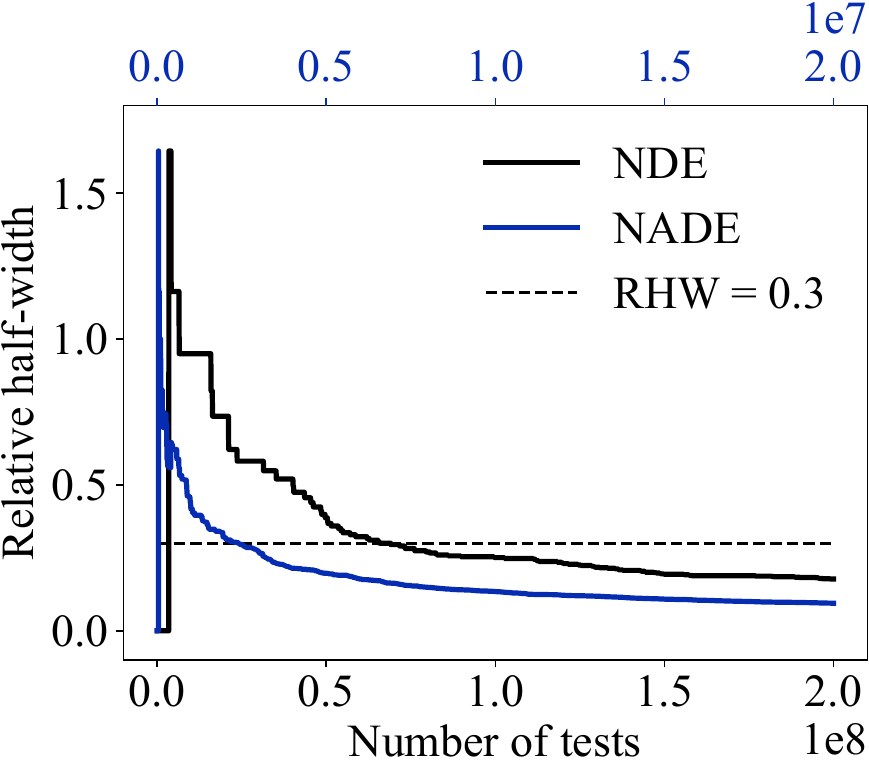}\label{subfig:RHW_NDE_NADE_AV_2}}
  \quad
  \subfigure[AV-III]{\includegraphics[width=\subfiglen]{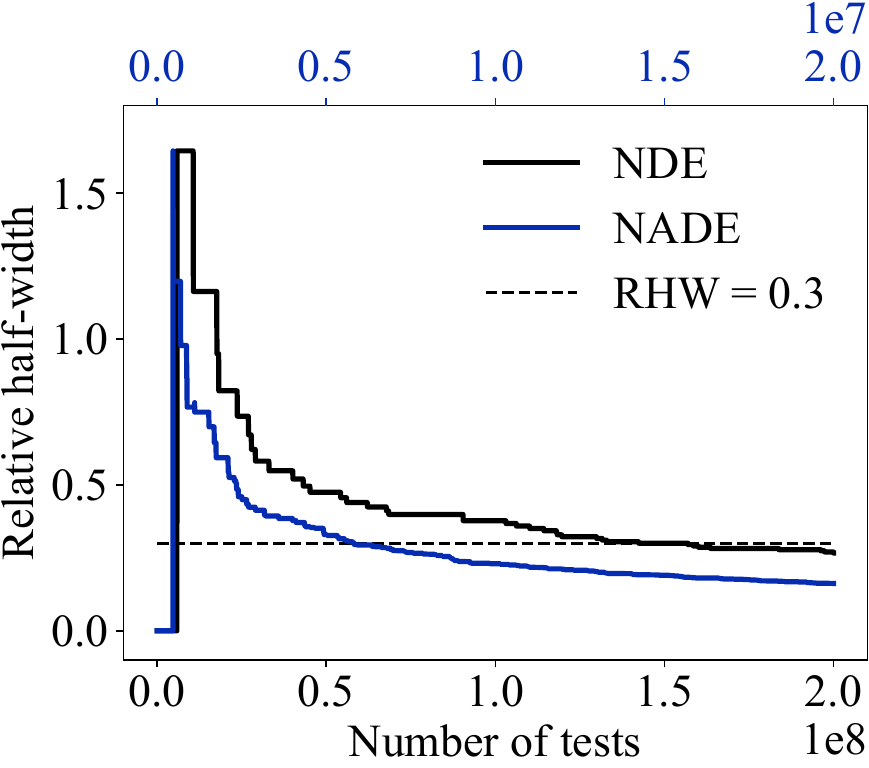}\label{subfig:RHW_NDE_NADE_AV_3}}
  \caption{The crash rate estimations for (a) AV-I, (b) AV-II and (c) AV-III in NDE and NADE, and corresponding RHW for (d) AV-I, (e) AV-II and (f) AV-III.}
  \label{fig:CR_RHW_NDE_NADE}
\end{figure*}

\begin{figure*}[!t]
  \centering
  \setlength{\subfiglen}{5cm}
  \subfigure[AV-I]{\includegraphics[width=\subfiglen]{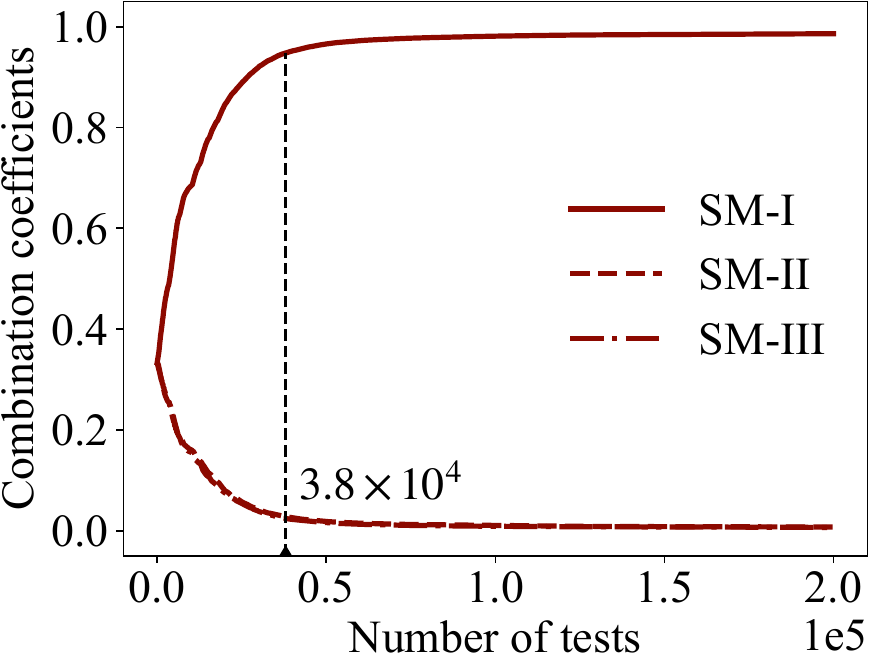}\label{subfig:combination_coefficients_AV_1_IDM}}
  \quad
  \subfigure[AV-II]{\includegraphics[width=\subfiglen]{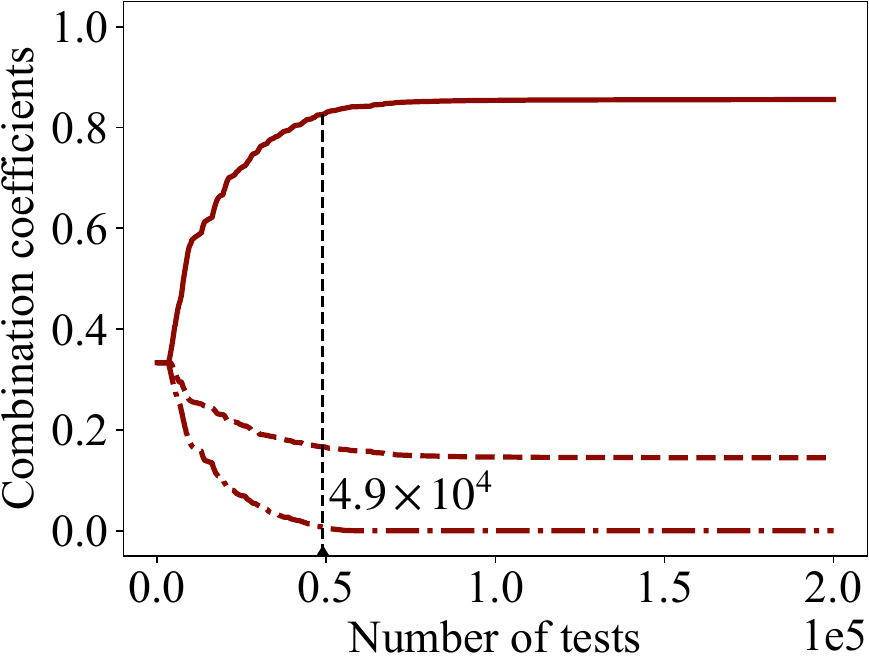}\label{subfig:combination_coefficients_AV_2_VT_IDM}}
  \quad
  \subfigure[AV-III]{\includegraphics[width=\subfiglen]{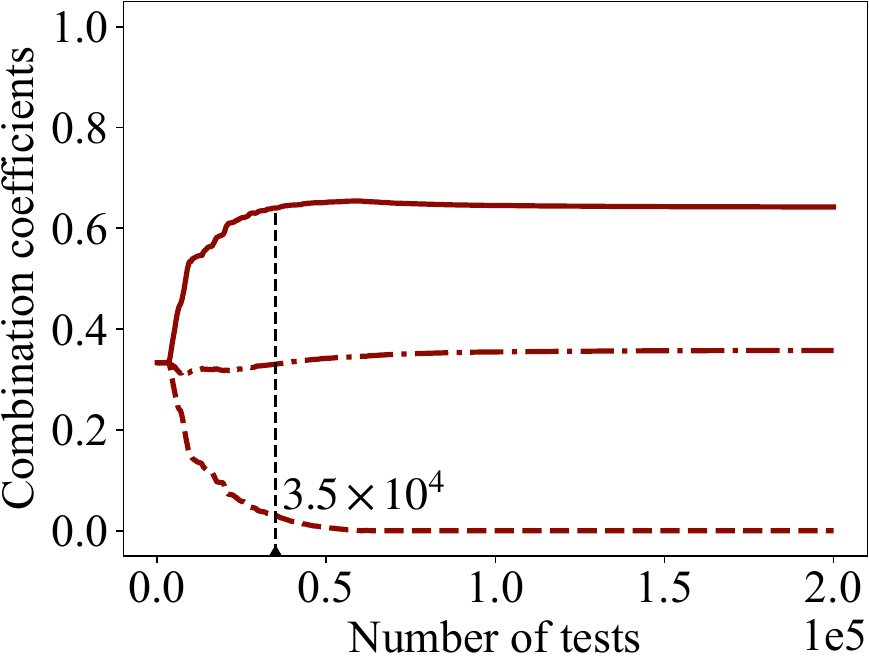}\label{subfig:combination_coefficients_AV_3_PPO}}
  \\
  \subfigure[AV-I]{\includegraphics[width=\subfiglen]{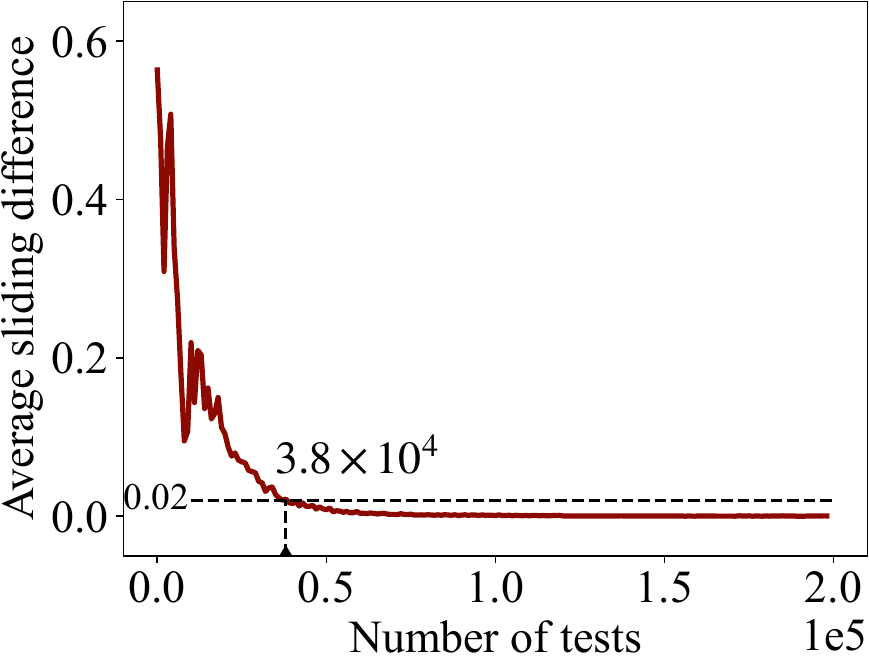}\label{subfig:average_sliding_diff_AV_1_IDM}}
  \quad
  \subfigure[AV-II]{\includegraphics[width=\subfiglen]{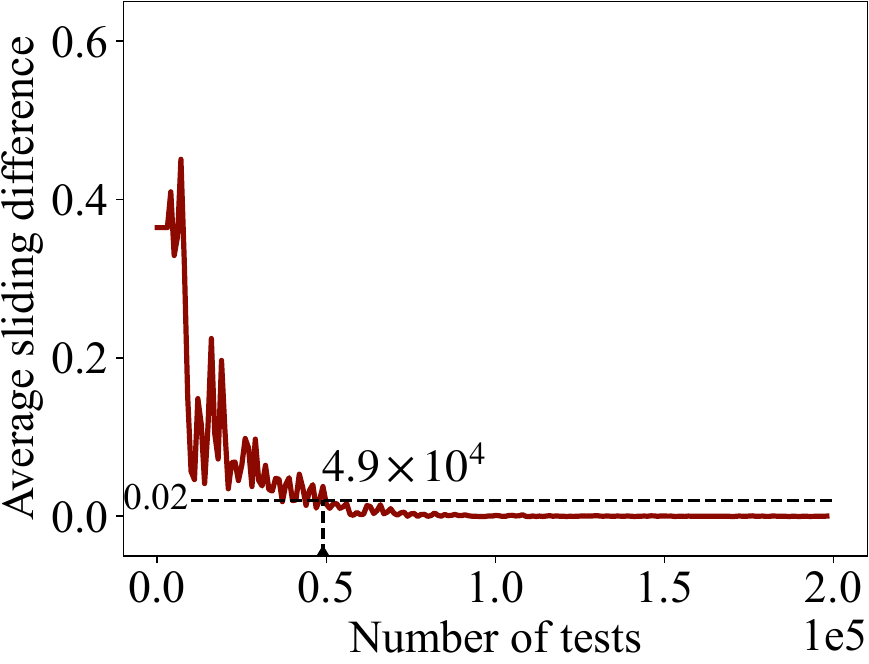}\label{subfig:average_sliding_diff_AV_2_VT_IDM}}
  \quad
  \subfigure[AV-III]{\includegraphics[width=\subfiglen]{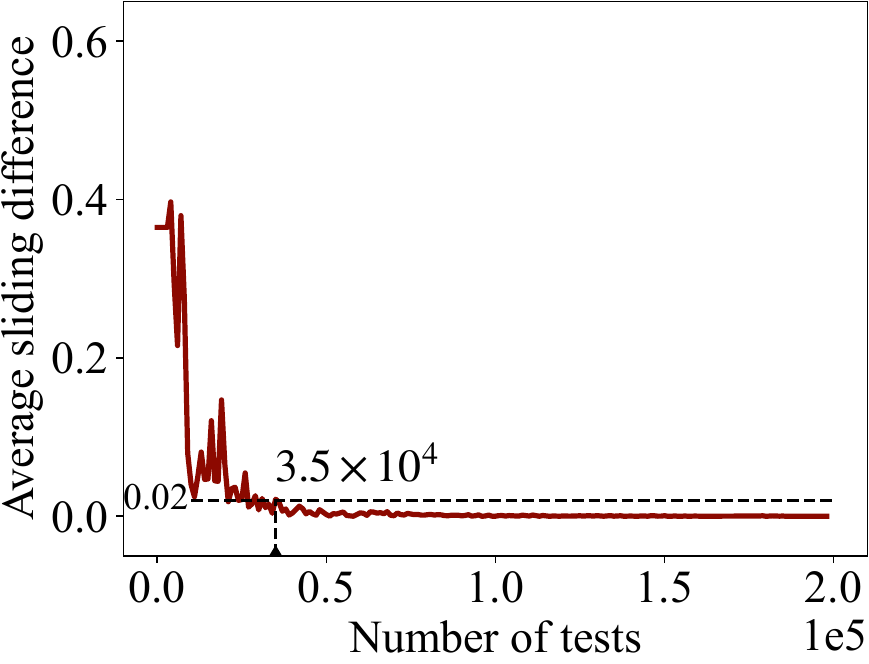}\label{subfig:average_sliding_diff_AV_3_PPO}}
  \caption{The combination coefficients optimized by DenseRL with the adaptive policy for (a) AV-I, (b) AV-II and (c) AV-III, and the corresponding ASD for (d) AV-I, (e) AV-II and (f) AV-III.}
  \label{fig:combination_coefficients_ASD}
\end{figure*}

\begin{figure*}[!t]
  \centering
  \setlength{\subfiglen}{5cm}
  \subfigure[AV-I]{\includegraphics[width=\subfiglen]{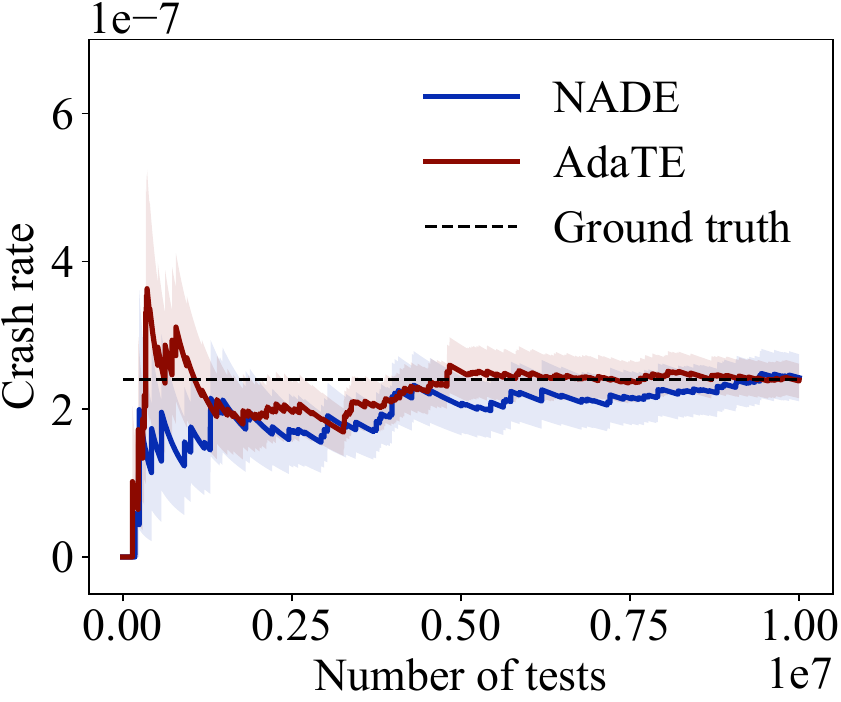}\label{subfig:CR_NADE_AdaTE_AV_1}}
  \quad
  \subfigure[AV-II]{\includegraphics[width=\subfiglen]{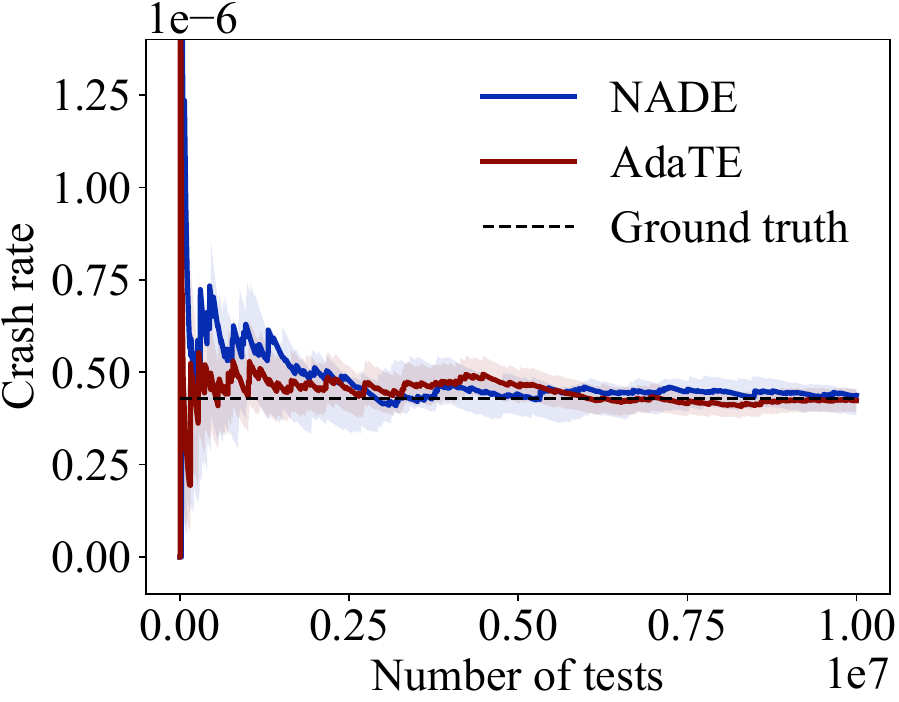}\label{subfig:CR_NADE_AdaTE_AV_2}}
  \quad
  \subfigure[AV-III]{\includegraphics[width=\subfiglen]{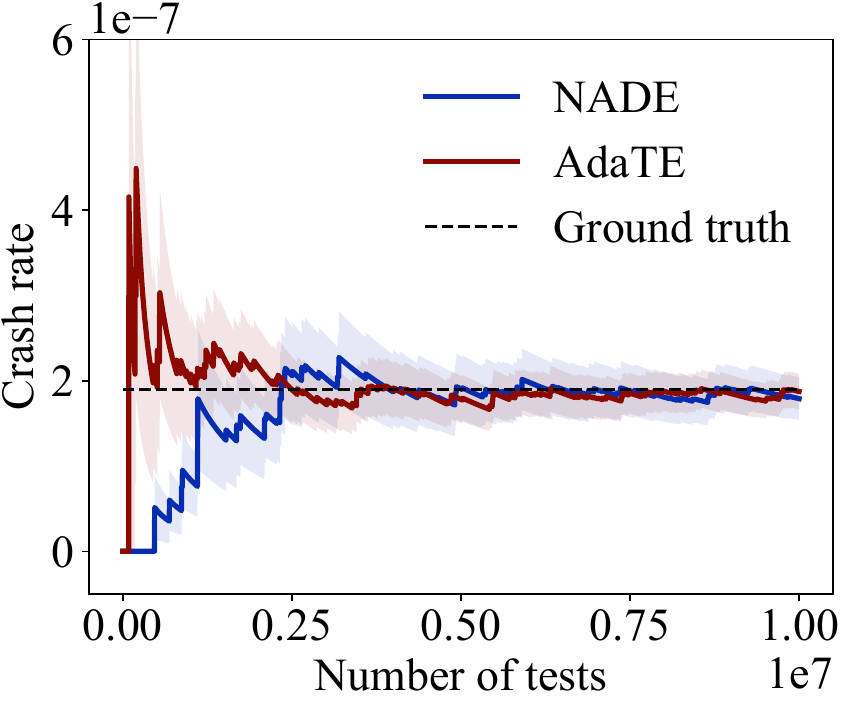}\label{subfig:CR_NADE_AdaTE_AV_3}}
  \\
  \subfigure[AV-I]{\includegraphics[width=\subfiglen]{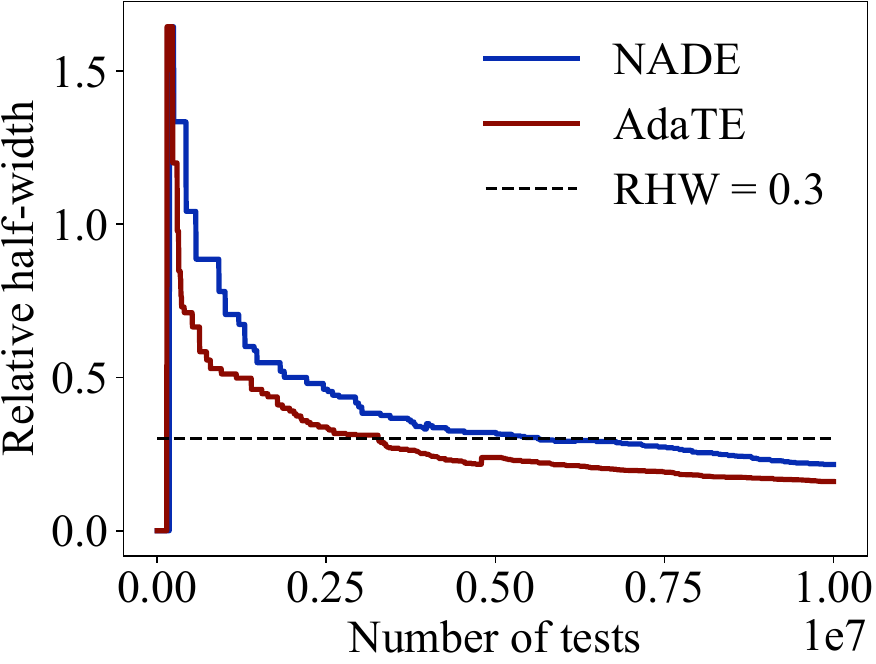}\label{subfig:RHW_NADE_AdaTE_AV_1}}
  \quad
  \subfigure[AV-II]{\includegraphics[width=\subfiglen]{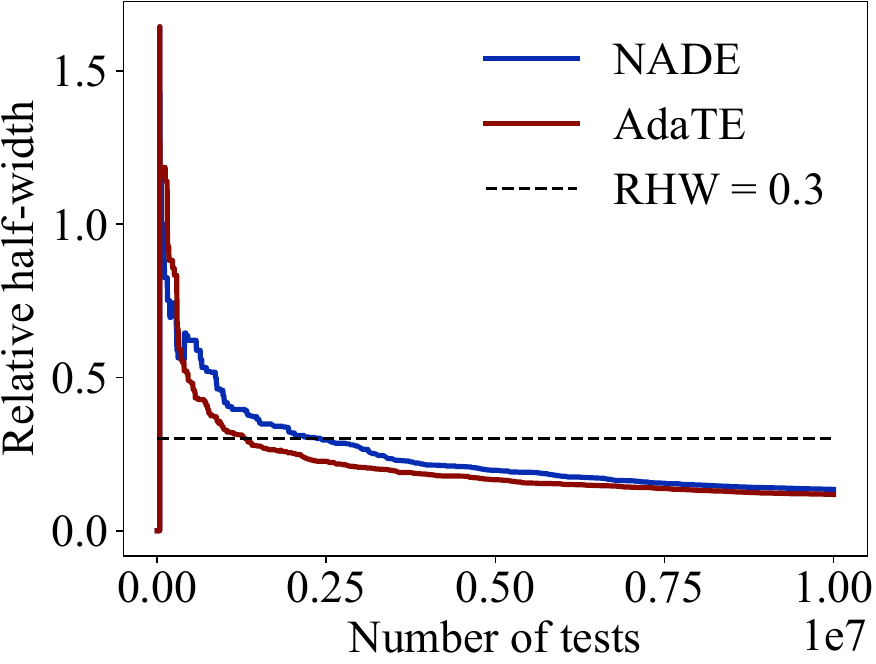}\label{subfig:RHW_NADE_AdaTE_AV_2}}
  \quad
  \subfigure[AV-III]{\includegraphics[width=\subfiglen]{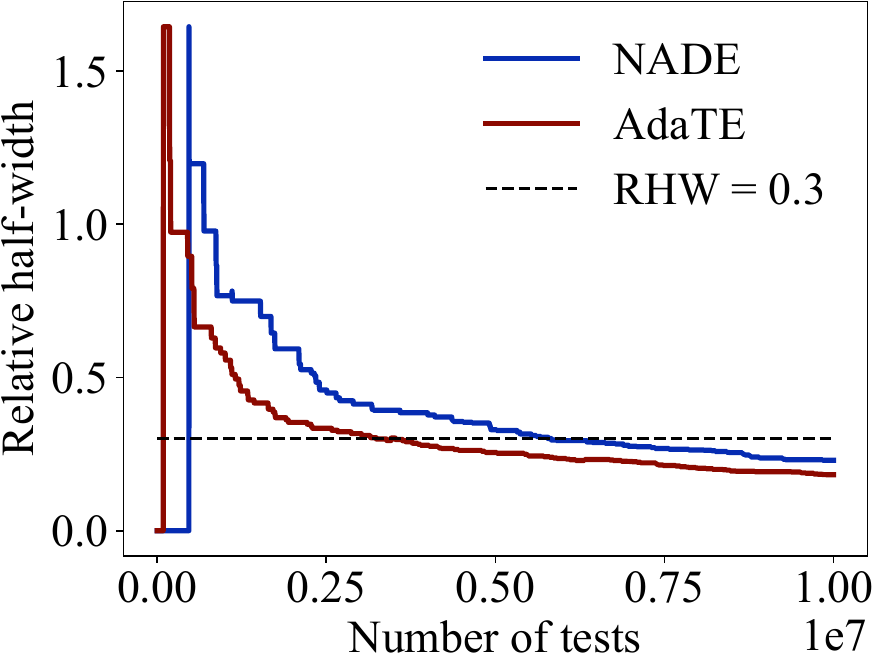}\label{subfig:RHW_NADE_AdaTE_AV_3}}
  \\
  \subfigure[AV-I]{\includegraphics[width=\subfiglen]{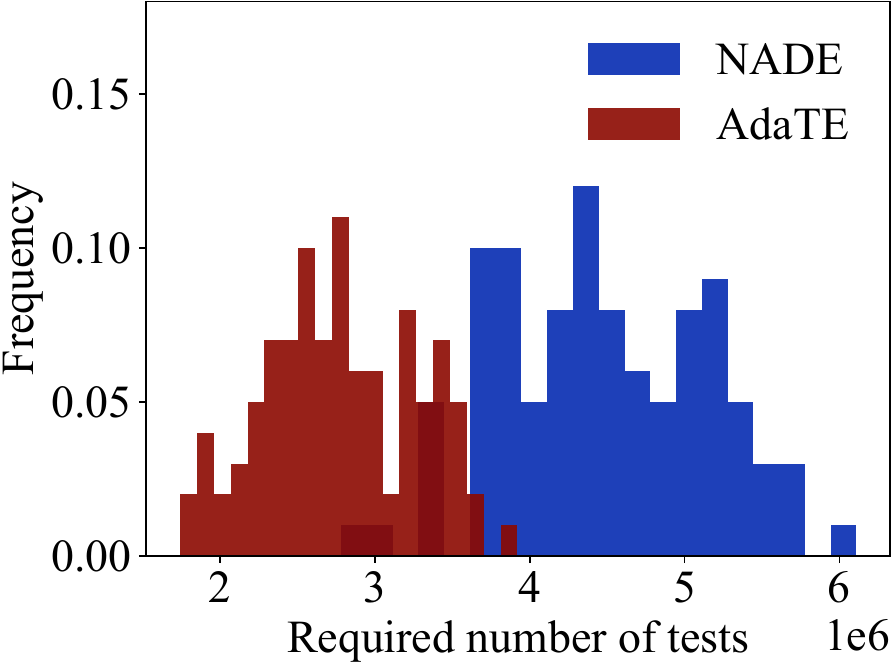}\label{subfig:RNoT_dist_NADE_AdaTE_AV_1}}
  \quad
  \subfigure[AV-II]{\includegraphics[width=\subfiglen]{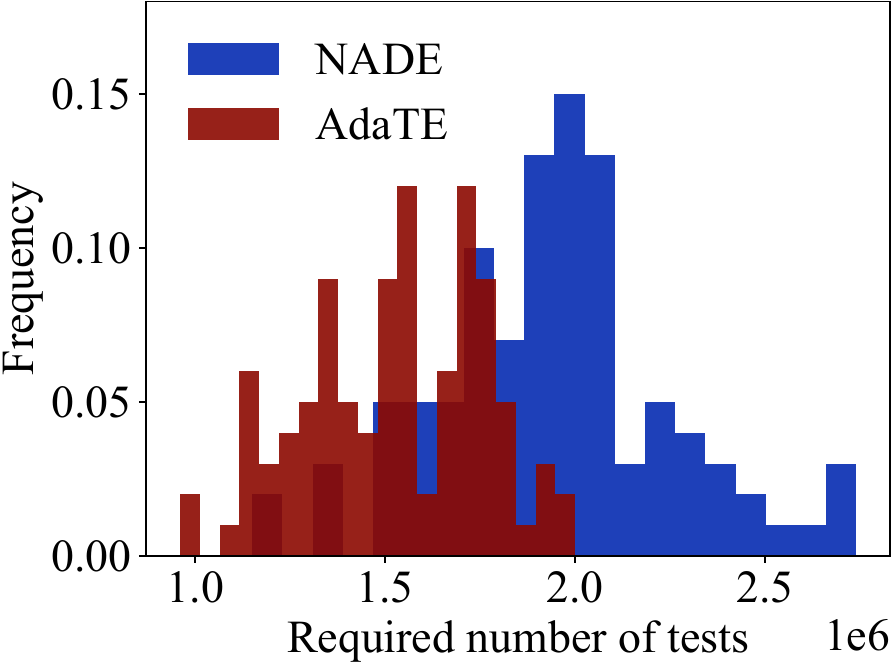}\label{subfig:RNoT_dist_NADE_AdaTE_AV_2}}
  \quad
  \subfigure[AV-III]{\includegraphics[width=\subfiglen]{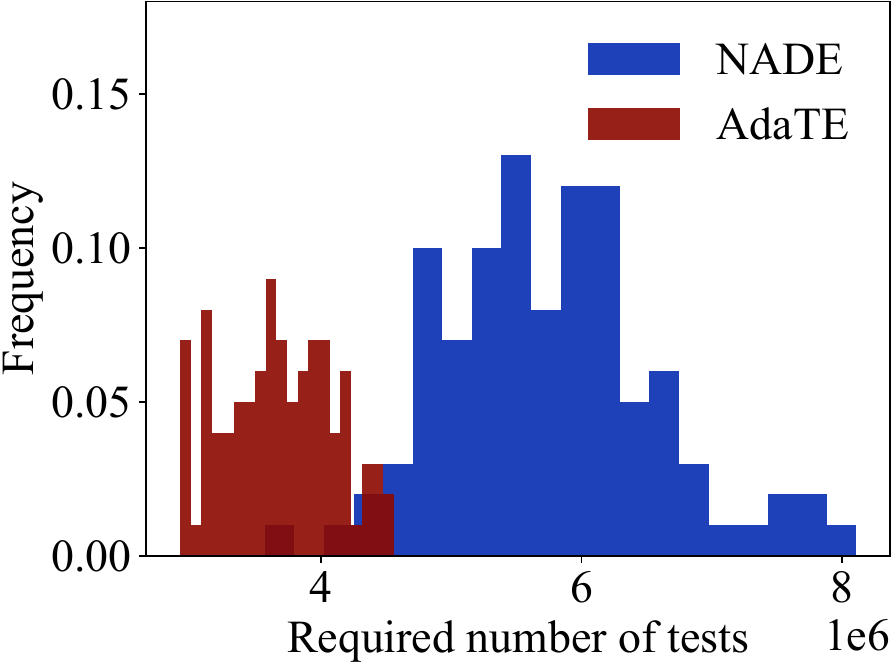}\label{subfig:RNoT_dist_NADE_AdaTE_AV_3}}
  \caption{The crash rate estimations for (a) AV-I, (b) AV-II and (c) AV-III in NADE and AdaTE, RHW of crash rate estimations for (d) AV-I, (e) AV-II and (f) AV-III, and frequency distributions of bootstrapped required number of tests for (g) AV-I, (h) AV-II and (i) AV-III.}
  \label{fig:CR_RHW_NADE_AdaTE}
\end{figure*}

To demonstrate the failure cases of NADE due to surrogate-to-real gaps, we test AV-I in the NADE where the importance function is constructed from SM-III. Fig.~\ref{subfig:CR_NDE_NADE_AV_1_SM_3} shows the crash rate estimated by NDE and NADE (SM-III), where the bottom $x$-axis represents the number of tests of NDE and the top $x$-axis stands for the number of tests of NADE (SM-III). The relative half-width (RHW) \cite{zhao2016accelerated} is used as a proxy to measure the convergence of crash rate estimation, which is shown in Fig.~\ref{subfig:RHW_NDE_NADE_AV_1_SM_3}. It can be seen that NADE  (SM-III) fails to converge to the ground truth crash rate estimated by NDE. As SM-III is more conservative than AV-I, the corresponding importance function may omit critical scenarios, leading to this failure.

To bolster the evaluation robustness of NADE, we use three SMs with average combination coefficients (i.e., $\bm{\alpha}=[1/3,1/3,1/3]^\top$) to establish the importance function. Fig.~\ref{fig:CR_RHW_NDE_NADE} shows the crash rate estimated in this new NADE and the corresponding RHW for AV-I, AV-II and AV-III, respectively. It can be found that for all three AVs, NADE converges to the same crash rate estimation as NDE, while using much less number of tests for reaching the 0.3 RHW threshold.

Using multiple SMs with average combination coefficients could improve evaluation robustness of NADE, but the evaluation efficiency may be compromised, as such NADE is not customized for any specific AV under test. We optimize the combination coefficients by DenseRL. Fig.~\ref{fig:combination_coefficients_ASD}(a)-(c) uncover that DenseRL is able to optimize the combination coefficients effectively and efficiently. In particular, the optimized combination coefficients for AV-I, AV-II and AV-III are $\bm{\alpha}_{\text{AV-I}}=[0.95, 0.03, 0.02]^\top$ (the ground truth is $\bm{\alpha}_{\text{AV-I}}^*=[1,0,0]^\top$), $\bm{\alpha}_{\text{AV-II}}=[0.82, 0.16, 0.02]^\top$, and $\bm{\alpha}_{\text{AV-III}}=[0.65, 0.02, 0.33]^\top$, respectively. To reach the ASD threshold (0.02), the required number of tests for AV-I, AV-II and AV-III are $3.8\times10^4$, $4.9\times10^4$, and $3.5\times10^4$, respectively, as shown in Fig.~\ref{fig:combination_coefficients_ASD}(d)-(f). Then the AdaTE can be generated by using three SMs with the optimized combination coefficients. 


\begin{table}[!t]
  \centering
  \setlength{\tabcolsep}{4pt}
  \renewcommand{\arraystretch}{1.3}
  \caption{Average required number of tests and average acceleration ratios for AV-I, AV-II and AV-III.}
  \label{tab:bootstrapped_evaluation_results}
  \begin{tabular}{c|lll}
    \hline
    Methods & AV-I (AAR) & AV-II (AAR) & AV-III (AAR) \\
    \hline
    NDE & $1.23\times10^8$ & $7.01\times10^7$ & $1.57\times10^8$ \\
    NADE & $4.46\times10^6$ (28) & $1.94\times10^6$ (36) & $5.74\times10^6$ (27) \\
    AdaTE & $2.78\times10^6$ (44) & $1.52\times10^6$ (46) & $3.67\times10^6$ (43) \\
    \hline 
  \end{tabular}
\end{table}

To investigate the performance of AdaTE, we compare its results with NADE, as shown in Fig.~\ref{fig:CR_RHW_NADE_AdaTE}. It can be seen from Fig.~\ref{fig:CR_RHW_NADE_AdaTE}(a)-(c) that AdaTE achieves the same crash rate estimation as NADE for all three AVs. To reach the 0.3 RHW threshold, AdaTE requires less number of tests than NADE, as shown in Fig.~\ref{fig:CR_RHW_NADE_AdaTE}(d)-(f). To alleviate the stochasticity of experiments, we bootstrap the testing results by shuffling 100 times. The frequency distributions of required number of tests for AV-I, AV-II and AV-III are shown in Fig.~\ref{fig:CR_RHW_NADE_AdaTE}(g)-(i), respectively. The average required number of tests and average acceleration ratios (AARs) of NDE, NADE and AdaTE for three AVs are shown in Table~\ref{tab:bootstrapped_evaluation_results}, where AARs (presented in parentheses) are ratios of the average required number of tests in NADE and AdaTE with respect to NDE. Compared with NADE, AdaTE can reduce 37.67\%, 21.64\%, 36.06\% number of tests for AV-I, AV-II and AV-III, respectively, revealing significant performance for increasing evaluation efficiency while enhancing evaluation robustness.

\section{Conclusion}
\label{sec:conclusion}

This paper proposes the dense reinforcement learning approach, which is designed to facilitate adaptive testing for a wide range of CAVs. The key idea involves learning exclusively the values associated with critical state-action pairs that exhibit significant surrogate-to-real gaps. By integrating DenseRL with an adaptive policy for determining the regression target and employing QP for the regression of combination coefficients, the AdaTE can be generated for diverse CAVs. The effectiveness of our method is validated in high-dimensional overtaking scenarios, revealing AdaTE's superior evaluation efficiency compared to both NDE and NADE. One limitation of this work is that we only consider discretized state and action spaces. Extending our approach to continuous cases warrants further exploration. Additionally, we have solely concentrated on the adaptive generation of testing scenarios, without delving into the adaptive evaluation of testing results. Future endeavors will aim to integrate both aspects.

\bibliographystyle{IEEEtran}
\bibliography{IEEEabrv,reference.bib}

\begin{IEEEbiography}[{\includegraphics[width=1in,height=1.25in,clip,keepaspectratio]{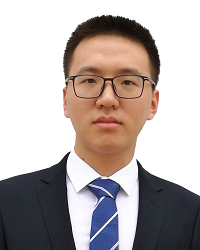}}]{Jingxuan Yang}
  received the bachelor's degree from the School of Mechanical Engineering and Automation, Harbin Institute of Technology, Shenzhen, China, in 2020. He is currently working towards the Ph.D. degree with the Department of Automation, Tsinghua University, Beijing, China. His current research interests include adaptive testing and evaluation of connected and automated vehicles.
\end{IEEEbiography}

\begin{IEEEbiography}[{\includegraphics[width=1in,height=1.25in,clip,keepaspectratio]{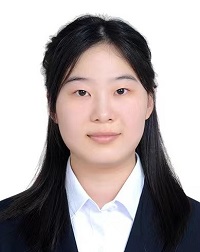}}]{Ruoxuan Bai}
  received the bachelor’s degree from the School of Mechanical and Vehicular Engineering, Beijing Institute of Technology, Beijing, China, in 2022. She is currently working towards the master's degree with the Department of Automation, Tsinghua University, Beijing, China. Her current research interests include testing and evaluation of intelligent systems.
\end{IEEEbiography}

\begin{IEEEbiography}[{\includegraphics[width=1in,height=1.25in,clip,keepaspectratio]{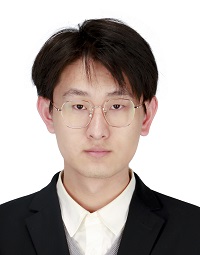}}]{Haoyuan Ji}
  is currently working towards the bachelor's degree with the Department of Automation, Tsinghua University, Beijing, China. His current research interests include adaptive testing, reinforcement learning and computer vision.
\end{IEEEbiography}

\begin{IEEEbiography}[{\includegraphics[width=1in,height=1.25in,clip,keepaspectratio]{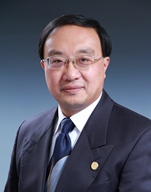}}]{Yi Zhang}
  (Senior Member, IEEE) received the B.S. and M.S. degrees from Tsinghua University, China, in 1986 and 1988, respectively, and the Ph.D. degree from the University of Strathclyde, U.K., in 1995. He is currently a Professor in control science and engineering with Tsinghua University. His current research interests focus on intelligent transportation systems. His active research areas include intelligent vehicle-infrastructure cooperative systems, analysis of urban transportation systems, urban road network management, traffic data fusion and dissemination, and urban traffic control and management. His research fields also cover the advanced control theory and applications, advanced detection and measurement, as well as systems engineering.
\end{IEEEbiography}

\begin{IEEEbiography}[{\includegraphics[width=1in,height=1.25in,clip,keepaspectratio]{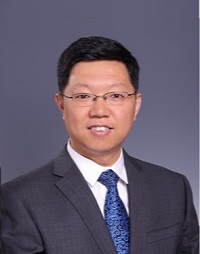}}]{Jianming Hu}
  (Senior Member, IEEE) received the B.E., M.E., and Ph.D. degrees in 1995, 1998, and 2001, respectively. He is currently an Associate Professor with the Department of Automation (DA), Tsinghua University. He has presided and participated in more than 20 research projects granted from the Ministry of Science and Technology of China, National Science Foundation of China, and other large companies with more than 30 journal articles and more than 100 conference papers. His recent research interests include networked traffic flow, large-scale traffic information processing, intelligent vehicle infrastructure cooperation systems (V2X or Connected Vehicles), and urban traffic signal control.
\end{IEEEbiography}

\begin{IEEEbiography}[{\includegraphics[width=1in,height=1.25in,clip,keepaspectratio]{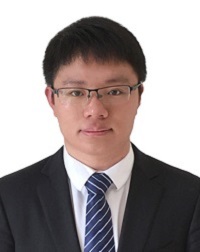}}]{Shuo Feng}
  (Member, IEEE) received the bachelor's and Ph.D. degrees in the Department of Automation at Tsinghua University, China, in 2014 and 2019, respectively. He was a postdoctoral research fellow in the Department of Civil and Environmental Engineering and also an Assistant Research Scientist at the University of Michigan Transportation Research Institute (UMTRI) at the University of Michigan, Ann Arbor. He is currently an Assistant Professor in the Department of Automation at Tsinghua University. His research interests lie in the development and validation of safety-critical machine learning, particularly for connected and automated vehicles. He was a recipient of the Best Ph.D. Dissertation Award from the IEEE Intelligent Transportation Systems Society in 2020 and the ITS Best Paper Award from the INFORMS TSL society in 2021. He is an Associate Editor of the \textsc{IEEE Transactions on Intelligent Vehicles} and an Academic Editor of the \textit{Automotive Innovation}.
\end{IEEEbiography}

\end{document}